\documentclass[11pt]{article}
\usepackage[left=1in, right=1in, top=1in, bottom=1in]{geometry}
\usepackage{latexsym}
\usepackage{graphicx}
\usepackage{mathptmx}
\usepackage[T1]{fontenc}
\usepackage{booktabs} 
\usepackage{tcolorbox} 
\usepackage{enumitem}
\usepackage{subcaption} 
\usepackage{amsmath}
\usepackage{amsfonts}
\usepackage{amssymb}
\usepackage{amsbsy}
\usepackage{amsthm}
\usepackage{wrapfig}
\usepackage{tikz}
\usepackage[numbers]{natbib}
\usepackage{fancyhdr}
\usepackage{hyperref}

\fancypagestyle{titlepagefooter}{
  \fancyhf{} 
  \fancyfoot[C]{\small An abridged version of this paper has been accepted at the \href{https://meetings.informs.org/wordpress/wsc2026/}{2026 Winter Simulation Conference (WSC 2026)}.}

}
\usetikzlibrary{arrows.meta, positioning, shapes.geometric}

\setlist[itemize]{leftmargin=*, nosep, labelsep=1em}
\setlist[enumerate]{leftmargin=*, nosep, labelsep=1em}
\newtheorem{definition}{Definition}

\newcommand{\prob}{\textsc{FlowInterdiction}}
\newcommand{\probnetwork}{\textsc{NetworkSynthesis}}
\newcommand{\probbinarynode}{\prob}

\newcommand{\probcontinuousnode}{\textsc{(Continuous, Node)}-\prob}
\newcommand{\probrobustbinarynode}{\textsc{Robust}-\prob}
\newtheorem{observation}{Observation}

\begin{document}

\title{Robust Network Flow Interdiction Problems with Applications to Counter-Narcotics}
\author{Diksha Gupta\textsuperscript{1}, Madhav V. Marathe \textsuperscript{1,2}, Anil Vullikanti \textsuperscript{1,2} \\
[11pt]
\textsuperscript{1} Biocomplexity Institute, University of Virginia, USA \\
\textsuperscript{2} Department of Computer Science, University of Virginia, USA \\
} 
\date{} 

\maketitle
\thispagestyle{titlepagefooter}

\begin{abstract}
\noindent Interdiction problems arise in a number of application areas, including  global security, supply chains, and critical infrastructure protection --- the goal is inhibit the movement of goods, people or information. An area of particular interest is counter-narcotics, where nodes or edges in a network are placed under surveillance or blocked to minimize the flow of illicit drugs from source to the destination. A fundamental challenge in this narco-traffic interdiction is data scarcity: available datasets are extremely limited by the very nature of the problem and provide only partial and uncertain views of trafficking networks. Thus it is of interest to develop interdiction methods that take this inherent lack of information into account by developing robust solutions.

In this paper we initiate the study of network flow interdiction problems under network uncertainty. First, using a limited real-world dataset, we generate an ensemble of plausible network realizations representing alternative trafficking scenarios. The method
combines simulations with mathematical programming techniques to generate network ensembles that are consistent with the observed data. Second, we formulate the robust network flow interdiction problem and develop a integer linear program to solve the problem. 
We evaluate the optimal interdiction strategy and obtain the residual flows over the scenarios. Our analysis reveals that  even modest budgets can yield significant flow reductions. However, optimal solutions vary substantially across scenarios, motivating the need for robust solutions. We show that the robust strategy achieves near-optimal performance across all near-real world realizations while remaining stable under structural uncertainty. This simulation-driven approach provides a principled basis for policy analysis and supports maximizing the return on interdiction investments in uncertain, data-limited environments.
\end{abstract}

\section{Introduction}\label{sec:intro}
A number of practical problems in the field of security, supply chain and infrastructure involve finding strategies to detect and restrict an adversary's actions so as to minimize their overall impact \cite{steinrauf1991network}. Such problems are often viewed as interdiction problems. Informally,  interdiction refers to actions that serve to block or otherwise inhibit an adversary’s operations, and often regards attacks against supply chain operations or communications. In game theory and operations research (OR), interdiction problems are often viewed as two person leader follower game: the leader interdicts to minimize the impact of followers action that is also viewed as an optimization problem. Here we focus on network interdiction problems, where the game is played on a network. Researchers also view this as a bi-level optimization problem -- leader minimizes the impact of the follower's maximization strategy. The field of network interdiction has been studied extensively over the last five decades due to their wide applicability; see \cite{wollmer1964removing,phillips1993network,burch2003decomposition,chestnut2017hardness,zenklusen2010network,pan2016interdiction,matuschke2017protection} for important publications and \cite{ausiello2026interdiction,smith2020survey} for recent surveys on this topic. 
In this report, we will focus on network flow interdiction problem --- informally we seek to find a strategy for the leader under budget constraints to reduce the amount of flow that a follower can send between a source-destination pair.

Our motivation to study this problem arose due to its applications in interdicting the movement of illicit drugs \cite{wollmer1964removing,phillips1993network,burch2003decomposition}. 
However, the formulation is general and applies to interdicting other kinds of flows, including, human trafficking and moving contraband animals and goods.
Trafficking of drugs and precursor chemicals often occurs through trade and transportation networks, including remote regions and highways. These have been common studied as a flow, e.g., \cite{steinrauf1991network,phillips1993network,wood1993deterministic}, (sent by an attacker) between a source $s$ and a sink $t$ on networks, denoted $G=(V, E)$, where $V$ denotes the set of nodes (e.g., transit points and cities) with $s, t\in V$, and $E$ denotes the set of edges (e.g., transportation links), with a capacity $cap(e)$ for each $e\in E$, which indicates the maximum flow through the edge.
Flows can model transfer of physical materials in the network, both in security applications, e.g., over military supply lines, and in civilian applications, such as discharge from polluting pipe systems.
More generally, network flows are also used to model opportunities for movement of attackers in the network; such models have been used to study the defense of critical resources, such as airports and transportation lines, e.g., \cite{pita2008deployed,tsai2009iris}.

Flow interdiction problems involve interdicting components of the network (where interdiction models reduction in the capacity of these edges/nodes or their removal), so that the maximum possible flow from $s$ to $t$ in the residual network is minimized.
Different kinds of interdiction models have been studied,
such as edges or nodes \cite{phillips1993network,burch2003decomposition,chestnut2017hardness} or flow paths \cite{matuschke2017protection}.  
Further, interdiction can be continuous (where the flow through selected edges or paths can be reduced fractionally \cite{phillips1993network,burch2003decomposition,matuschke2017protection}) or
discrete \cite{chestnut2017hardness,pan2016interdiction} (where nodes or edges are completely removed from the network). 
While most of the work on flow interdiction has been on specific networks, there has been recognition of the uncertainty in such applications.

These problems are computationally very challenging.
The simplest flow interdiction problem is weakly NP-complete even for planar graphs and strongly NP-complete for general graphs  \cite{phillips1993network}, and hard to approximate within a factor of $n^{\epsilon}$ for any $\epsilon>0$ \cite{chestnut2017hardness}.
We briefly mention some of the main results here, and refer to \cite{ausiello2026interdiction,smith2020survey} for more details. \cite{phillips1993network,pan2016interdiction} developed algorithms for planar graphs.
The first polynomial time algorithm for general graphs was by Burch et al. \cite{burch2003decomposition}, who  design a bicriteria approximation.
The key idea is to solve a linear programming (LP) relaxation of the problem, and then find two feasible integral solutions whose convex composition equals the LP optimal solution.
There has also been a lot of work on new mixed integer programming (MIP) techniques for solving such problems, e.g., \cite{ausiello2026interdiction,ghare1971optimal,royset2007solving}.

A fundamental challenge in this application domain is extreme data sparsity. The underlying trafficking network, edge capacities, and route-level flows are typically unobserved, while available datasets provide only partial, noisy, and geographically aggregated measurements of successful interdiction events~\cite{mcsweeney2020reliable}. Consequently, network reconstruction is highly nuanced: many structurally distinct networks may be consistent with the same aggregate observations. Prior work has therefore relied on constructing realistic synthetic networks to enable analysis; an important paper in this direction is the work of~\cite{magliocca2019modeling}. We build on their work by developing an ensemble of networks that captures uncertainty in possible trafficking routes, capacities, and flow patterns. Rather than viewing this as the recovery of a single underlying network, \emph{we adopt a simulation-based robust optimization perspective that explicitly models structural uncertainty and seeks interdiction decisions that perform well across plausible network realizations.}

\subsection{Our Contributions and Overall Framework.}
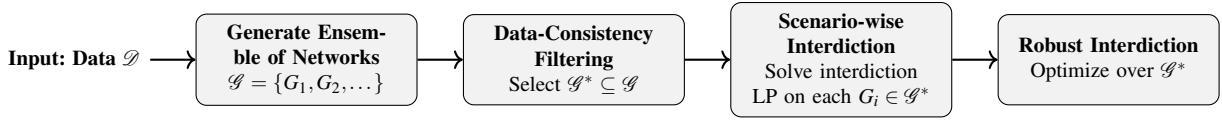
\begin{figure*}[t]
\centering
\begin{tikzpicture}[
node distance=0.6cm,
every node/.style={font=\scriptsize},
box/.style={
rectangle, draw, rounded corners,
fill=gray!10,
text width=0.16\textwidth,
minimum height=1.2cm,
inner sep=4pt,
align=center
},
arrow/.style={->, thick}
]

\node(data) {\textbf{Input: Data $\mathcal{D}$}};

\node[box, right=of data] (ensemble) {
\textbf{Generate Ensemble of Networks} \\
$\mathcal{G} = \{G_1,G_2,\dots\}$
};

\node[box, right=of ensemble] (filter) {
\textbf{Data-Consistency Filtering} \\
Select $\mathcal{G}^* \subseteq \mathcal{G} $
};

\node[box, right=of filter] (opt) {
\textbf{Scenario-wise Interdiction} \\
Solve interdiction LP on each $G_i \in \mathcal{G}^*$
};

\node[box, right=of opt] (robust) {
\textbf{Robust Interdiction} \\
Optimize over $\mathcal{G}^*$
};

\draw[arrow] (data) -- (ensemble);
\draw[arrow] (ensemble) -- (filter);
\draw[arrow] (filter) -- (opt);
\draw[arrow] (opt) -- (robust);

\end{tikzpicture}

\caption{\textbf{Summary of our interdiction pipeline under network uncertainty.}}
\label{fig:pipeline_horizontal}
\end{figure*}

We initiate the systematic study of robust network flow interdiction problems motivated by applications in narco-trafficking.
The overall framework is shown in Figure~\ref{fig:pipeline_horizontal}.
Our work builds on the earlier literature and extends it in two important ways: ($i$) methods to develop an ensemble of networks to overcome the inherent uncertainty in the network specification and 
($ii$) robust optimization formulation and analysis flow interdiction on network ensembles. Specific contributions include:

\begin{enumerate}
    \item \textbf{Synthesizing and analyzing ensemble of narco-traffic networks.} We introduce a data-driven framework that combines simulation with mathematical programming to generate an ensemble of edge-weighted capacity networks consistent with partial observation data (e.g., flows in some regions from the CCDB dataset for counter-narcotics, as discussed in Section \ref{sec:framework}). We refer to this as the \probnetwork{} problem, and develop a novel approach using techniques from network science and linear programming; by incorporating controlled stochastic perturbations in edges and capacities, our approach produces a distribution of plausible trafficking scenarios, enabling repeated evaluation of flows and interdiction outcomes under uncertainty.  The networks we generate match flows in CCDB within a small error bound. We observe that structural properties of these networks are generally well concentrated.

    \smallskip

    \item \textbf{Design, implementation and analysis of robust interdiction strategies.} We formalize interdiction as a scenario-based simulation problem, where single-shot interdiction decisions are evaluated across multiple network realizations. We define a quality metric to identify scenarios consistent with real-world data, allowing principled filtering of the ensemble and enabling meaningful comparison of strategies across heterogeneous but data-consistent networks. We then propose a robust optimization formulation that selects a single interdiction strategy performing well across all high-quality scenarios. This formulation captures the tradeoff between optimality and stability, and provides a systematic way to hedge against uncertainty in network structure. By explicitly modeling network uncertainty and evaluating strategies over an ensemble, our framework enables decision-makers to assess the return on investment of interdiction efforts. In particular, we quantify the tradeoff between performance and robustness, providing actionable guidance for resource allocation in data-limited counter-narcotics settings. Our robust strategy performs near-optimal, using which we were able to identify a core set of nodes that consistently reduce network capacity across interdiction budgets. 
\end{enumerate}

\medskip 

\noindent \textbf{Organization.} In Section~\ref{sec:prelim}, we provide basic definitions of the problem studied. Section~\ref{sec:framework} describes our method to create an ensemble of networks and carries out detailed structural analysis. Finally Section~\ref{sec:node_interdiction} provides a Integer linear programming formulation for the (robust) interdiction problems and carries out detailed computational experiments to derive policy level insights.

\section{Preliminaries}\label{sec:prelim}

Let $G=(V, E)$ denote a network. The set $V$ of nodes represents physical locations, such as cities, terminals, and specific regions, while an edge $e=(u, v)\in E$ represents a connection between nodes $u$ and $v$, e.g., a road link or a conduit for drug flows. Nodes $s, t\in V$ represent terminals; we assume flow moves from $s$ to $t$. We use $cap(e)$ to denote the capacity of an edge $e\in E$, which represents the maximum amount of flow through edge $e$. A flow is a function $f:E\rightarrow \mathbb{R}_{\geq 0}$ that satisfies $f(e)\leq cap(e)$ for all $e\in E$, along with flow conservation constraints at all nodes $v\in V$, $v\neq s,t$. Let $f^{max}(G, s, t)$ denote the maximum flow from $s$ to $t$ in $G$.

From prior work on counter-narcotics, it is clear that the networks are generally not known, at least not fully, and need to be synthesized from available data. For a set $\mathcal{R}\subset 2^V$ of node subsets, where each $R\in\mathcal{R}$ is a subset of nodes in $V$ corresponding to a geographical region, we assume that estimates of the total flow $g(R)$ through nodes within $R$ are known. This corresponds to the data available for counter-narcotics, as discussed in Section~\ref{sec:framework}. We consider the network synthesis problem, where the set $V$ of nodes is known, typically based on regions identified in prior literature, e.g., \cite{magliocca2019modeling}, but the edges and capacities need to be inferred. We say that the inferred network is ``consistent'' with observed estimates $g(\cdot)$ if there is a flow $f(\cdot)$ such that the induced regional flow through each $R\in\mathcal{R}$ approximates $g(R)$ within a prescribed tolerance. We define the notion of consistency in Section~\ref{sec:est_flow}.

\begin{definition}
    \textbf{Network Synthesis problem} (\probnetwork). Given a set $V$ of nodes, terminals $s, t\in V$, and total flow estimates $g(R)$ for each $R\in\mathcal{R}$, infer the distribution of networks $G(V, E)$ and capacities $cap(e), e\in E$, such that it has a maximum flow $f(\cdot)$ consistent with the estimates $g(\cdot)$.
\end{definition}

Note that there can be an ensemble of networks consistent with the observations; therefore, our goal is to infer the distribution of such networks, and examine properties over the distribution.

Let $c(v)$ and $c(e)$ denote the cost of interdicting node $v\in V$ and edge $e\in E$, respectively; this could model monitoring the node or edge (e.g., through checkpoints) so that any illicit flow is disrupted.
For a subset of nodes $S\subset V$, we define $G[V\setminus S]$ as the graph induced after nodes in $S$ are removed. Similarly, for a subset of edges $S'\subset E$, we define $G[E\setminus S']$ as the graph induced after edges in $S'$ are removed. If a subset $S\subset V$ of nodes is interdicted, the maximum flow in the residual network $G[V \setminus S]$ is $f^{max}(G[V\setminus S], s, t)$; similarly, the residual flow $f^{max}(G[E\setminus S'], s, t)$ after edge interdiction of subset $S'$ is defined.

The node interdiction problem involves choosing the best subset of nodes $S$ with $\sum_{v\in S} c(v)\leq B$ within a budget constraint $B$, which minimizes the maximum flow in the residual network, e.g., \cite{phillips1993network,burch2003decomposition}. In this formulation, when a node is interdicted, all the flow  incident on the edges is reduced to 0 -- in other words, when a node is interdicted, one can assume that the node and the edges incident on the node are deleted from the network. We will call this a \emph{0/1-version} (or \emph{binary}) of the interdiction problem. This problem is defined formally below.

\begin{definition}
\textbf{Node-based optimal network flow interdiction problem} (\probbinarynode).
Given a flow network $G=(V, E)$, terminals $s, t$, capacity $cap(\cdot)$, cost $c(\cdot)$, and budget $B$, choose a subset $S\subset V$ such that $c(S)=\sum_{v\in S} c(v)\leq B$ and $f^{max}(G[V\setminus S], s, t)$ is minimized.
\end{definition}

In our setting we have an ensemble of networks $G_i=(V, E_i), i=1,\ldots, N$, and an interdiction solution $S_i$ for $G_i$ need not be effective for a different network $G_{i'}$.
We study a robust version of the flow interdiction problem, where the goal is to find a solution $S$ whose interdiction minimizes the maximum residual flow in each network; other notions of robustness can also be considered. We define the \textbf{robust version} of the node interdiction problem below.

\begin{definition}
    \textbf{Node-based optimal robust network flow interdiction problem} (\probrobustbinarynode).
    Given a set of flow networks $G_i(V, E_i), i=1,\ldots,N$, terminals $s, t$, capacity $cap(\cdot)$, cost $c(\cdot)$, and budget $B$, choose a subset $S\subset V$ such that $c(S)=\sum_{v\in S} c(v)\leq B$ and $\max_i f^{max}(G_i[V\setminus S], s, t)$ is minimized.
\end{definition}

Note that even the simplest node-based interdiction problems are computationally intractable, e.g., \cite{phillips1993network,chestnut2017hardness}.

\section{Generation \& Analysis of Ensemble of Networks}\label{sec:framework}
A key challenge in studying interdiction problems is the limited availability of reliable data on illicit trafficking flows. By their nature, such activities are only partially observed, and the underlying trafficking networks are rarely fully known. Consequently, both the network structure, defined by nodes $V$ and edges $E$, and the associated edge capacities are inherently uncertain. 
Edge capacities, which represent the volume of illicit materials trafficked through the network, are typically unobserved. 
Typically, only aggregate data on flows in some regions is available, e.g., the CCDB dataset discussed below.
While node interdiction costs are more readily estimated from the defender’s perspective, they may still exhibit significant variability due to differences in infrastructure, geography, and enforcement resources across regions.
Here, we develop an approach for the \probnetwork{} problem using data on partial observations.

\medskip 

\noindent \textbf{Datasets.}  We demonstrate our framework using a drug trafficking network in Latin America constructed from two complementary data sources.
\smallskip
\begin{itemize}
    \item \textbf{Narcologic Network.} We use the Narcologic network from~\cite{magliocca2019modeling,magliocca2021comparative} as the baseline topology. This directed graph consists of 156 nodes and 1006 edges representing potential trafficking routes across Central America. Nodes correspond to geographic locations within Guatemala, Honduras, El Salvador, Nicaragua, Costa Rica, and Panama. A single source node represents Colombia (origin), and a sink node represents Mexico (destination). 
The network consists of edges from a node to the origin or destination, and some random edges.
The network encodes directional flow toward the destination, along with additional downstream connections that decrease in number as routes approach the sink.

    \smallskip

    \item \textbf{CCDB Dataset.} The CCDB dataset provides temporal estimates of narcotic trafficking volumes aggregated at 22 administrative regions across Central America, covering Panama, Costa Rica, Nicaragua, El Salvador, Honduras, and Guatemala. 
\end{itemize}

\begin{table}[b!]
\centering
\scriptsize
\vspace{10pt}
\begin{tabular*}{\textwidth}{@{\extracolsep{\fill}}lll}
\toprule
\textbf{Parameter} & \textbf{Description} & \textbf{Values} \\
\midrule
$\tau_p$ & Edge-length pruning threshold used for sparsification  & $500,600,\dots,2000$ km \\
$\tau_d$ & Minimum Connectivity threshold as fraction of the average degree of G & $0.1, 0.2,..., 1.0$ \\
$\alpha$ & Scaling factor for edge-level minimum flows $m_{uv}$ & $0.001,0.01,0.1,1.0$  \\
$\delta$ & Perturbation radius controlling edge minimum flows & $0.0001,0.001,0.01,0.1$  \\
\bottomrule
\end{tabular*}
\caption{Ensemble of networks generation parameters. }
\label{tab:network_generation_parameters}
\vspace{-20pt}
\end{table}

\smallskip
\noindent
\textbf{Gaps.} \ While the nodes in the Narcologic network are based on prior literature on counter-narcotics, there is a lack of compelling justification for edges. Furthermore, other components needed for the flow interdiction problem, namely capacities and costs, are not available.

To address these challenges, we develop a simulation-based network generation framework for \probnetwork{} that constructs an ensemble of plausible trafficking networks consistent with partially observed data. The framework operates in two stages. First, we generate candidate network topologies that reflect structural uncertainty. Second, we estimate edge-level flows that are consistent with observed regional trafficking volumes. Section~\ref{sec:est_cap} describes the topology generation procedure, and Section~\ref{sec:est_flow} presents the flow estimation method. The parameters governing this process are summarized in Table~\ref{tab:network_generation_parameters}.

\begin{figure}[t!]
    \begin{tcolorbox}[colback=gray!8, colframe=gray!50, boxrule=0.6pt, left=6pt, right=6pt, top=6pt, bottom=6pt]
    \scriptsize

    \textbf{Generation of Ensemble Networks for Node Interdiction}

\smallskip

\textbf{Input:} Baseline network graph $G=(V,E)$ with node locations and edge set; regional flow volumes $g(R)$; Set of parameter configurations $\Theta$ consisting of pruning threshold $\tau_p$, fractional connectivity threshold $\tau_d$, scaling factor $\alpha$ and perturbation radius $\delta$.

\smallskip

\textbf{Output:} Ensemble of generated trafficking networks
$\{(G_1,\bar{f}_1),\dots,(G_{|\Theta|},\bar{f}_{|\Theta|})\}$.

\medskip

\textbf{Monte Carlo simulation loop.} For each parameter configuration $(\tau_p^i,\tau_d^i,\alpha^i,\delta^i) \in \Theta$:

\smallskip

    \begin{enumerate}

        \item \textbf{Generate the network graph.} Starting from the baseline graph $G=(V,E)$:
    
        \begin{enumerate}
            \item \emph{Estimate missing regional volumes.}  For each region $R$ corresponding to nodes without observed flow volume, estimate its flow $\hat{g}(R)$ as the mean of neighboring regions with known volumes. This process is applied iteratively until all regions are assigned a value. Regions that remain isolated are assigned the country-wide average volume.
        
            \item \emph{Edge sparsification and connectivity repair.}  Remove edges whose length exceeds $\tau_p^i$, and reintroduce them with probability inversely proportional to their length. Add edges between closest nodes until every node is reachable from the source and the sink is reachable from every node. Finally, for nodes with degree less than $\tau_d^i$ times average degree of G, introduce edges to nearby nodes until the degree threshold is satisfied. Let $E_i$ denote the resulting edge set.
        \end{enumerate}

        \smallskip
    
        \item \textbf{Generate the network flow capacities.} For the graph $G_i=(V,E_i)$ and estimated regional flows $\hat{g}(R)$:
    
        \begin{enumerate}
            \item \emph{Estimate edge-level minimum flows.}  For each region $R$ containing trafficking nodes, compute the average capacity per incident edge, denoted by $AvgCap(R)$, and assign a node-level minimum flow $\mu_u = AvgCap(R)\cdot \deg(u)$. For each edge $(u,v)$, compute a lower bound flow $m_{uv} = \alpha^i \min(\mu_u,\mu_v)$ and sample the edge minimum flow $\ell_{uv} \sim \mathrm{Uniform}[(1-\delta^i)m_{uv},(1+\delta^i)m_{uv}].$
        
            \item \emph{Estimate edge-flow capacities satisfying regional constraints.}  Solve the Network Flow LP from Figure~\ref{fig:network_flows} using the generated edge minimum flows and regional volume constraints. The final generated edge-flow capacities $\bar{f}_i(\cdot)$ are selected as the solution minimizing the consistency parameter $\epsilon$.
        \end{enumerate}

    \end{enumerate}
\end{tcolorbox}

    \caption{Procedure for generation of ensemble of networks including estimating edge capacities using partially observed data.}
    \label{fig:network_syn}
\end{figure}

\subsection{Synthesizing the topology of trafficking networks}
\label{sec:est_cap}

Our network generation procedure takes as input a baseline network consisting of a set of nodes with geographic locations and a corresponding edge set, and regional flow volumes. Recall from Section~\ref{sec:prelim} that each $R \in \mathcal{R}$ denotes a geographic region represented as a subset of nodes in the network. In our empirical setting, the baseline graph is the Narcologic network, and each region $R$ corresponds to an ADM1 administrative region (state or province) in Central America represented in the CCDB dataset or the Narcologic network. The quantity $g(R)$ denotes the trafficking-volume associated with region $R$. The CCDB dataset provides regional flow volumes for ADM1 regions corresponding to a subset of nodes in the Narcologic network, while the missing regional volumes for remaining nodes are estimated using the procedure described below. The complete list of administrative regions, together with their observed or estimated trafficking volumes, is provided in Appendix Table~\ref{tab:adm1_volume_estimates}.

The procedure consists of two main components: (i) estimating missing regional trafficking volumes, and (ii) constructing a topology that balances empirical realism with structural variability.

\medskip

\noindent \textbf{\emph{Estimate missing regional volumes.}}
We estimate the trafficking volumes for the administrative regions in the Narcologic network missing from the CCDB dataset. Let $\mathcal{C}_r$ denote the set of regions with known volumes, and for each such $R \in \mathcal{C}_r$, let $Cap(R)$ be the volume of trafficking observed through $R$ in CCDB. Then, for each missing region $R \in \mathcal{R}$ in the Narcologic network, we estimate its regional capacity iteratively. 

At iteration $t$, let $N(R)$ denote the set of neighboring administrative regions of $R$. The missing regional capacity is updated using available values from neighboring regions as:
$$ Cap^{(t+1)}(R)
= \frac{1}{\left|N(R) \cap \mathcal{C}_r^{(t)}\right|}
\sum_{R' \in N(R) \cap \mathcal{C}_r^{(t)}} Cap^{(t)}(R'),
$$
where $\mathcal{C}_r^{(t)}$ denotes the set of regions with available capacity estimates at iteration $t$, and the values for $R \in \mathcal{C}_r$ are kept fixed throughout the iterations. For the isolated regions, we assign country averages as the estimated volume.

\medskip

\noindent \textbf{\emph{Edge sparsification and connectivity repair.}} The Narcologic edge set in $G$ contains long-distance links directly connecting nodes to the source and sink, without empirical evidence supporting the feasibility of such routes. To improve structural realism under uncertainty, we generate an ensemble of edge sets by incorporating subsets of long-distance connections while preserving global connectivity to maintain end-to-end reachability. We describe our modeling details below.

\smallskip

\noindent For each parameter configuration $(\tau_p^i,\tau_d^i,\alpha^i,\delta^i) \in \Theta$, the resulting edge set $E_i$ is generated as follows:

\begin{enumerate}

\item \textbf{Pruning long edges with probabilistic reinsertion.} The edge-pruning threshold, $\tau_p$ controls the removal of long-distance connections in the baseline Narcologic network. We consider values between 500 km and 2000 km, spanning the range of routes in the edge set of Narcologic networks. Since reliable empirical information on the feasibility of individual trafficking routes is unavailable, we do not calibrate a single value of $\tau_p$. Instead, we treat it as a parameter and generate an ensemble of network realizations spanning this range.

\smallskip

For the geodesic distance threshold $\tau_p^i$, all edges with length exceeding $\tau_p^i$ are removed from $E$, yielding a sparsified edge set $\hat{E}_i$. Note that deterministic removal of all edges longer than $\tau_p^i$ can unrealistically eliminate plausible long-range trafficking routes and substantially reduce structural variability across generated networks. Hence, we reintroduce  each removed edge $e \in E \setminus \hat{E}_i$ with probability
$$ p = 1 - \frac{d(e)}{\max_{e' \in \hat{E}_i} d(e')},$$
where $d(e)$ denotes the geodesic length of edge $e$. This induces a distance-dependent decay, ensuring that longer routes are less likely to appear while still allowing occasional long-range connections, thereby preserving structural variability across generated networks. We provide a sensitivity analysis in Figure~\ref{fig:ablation_network_generation}.

\smallskip

\item \textbf{Connectivity repair.} To ensure feasibility of flow, we require that every node is reachable from the source and that the sink is reachable from every node. For nodes that violate this condition in $\hat{G}_i = (V, \hat{E}_i)$, we iteratively add edges to their nearest neighbors within a radius $r$ until reachability is restored. The radius $r$ is chosen minimally to satisfy this constraint. Additionally, because aggressive edge pruning can produce unrealistically sparse local structures, we enforce a minimum degree condition. This step reflects the role of geographic proximity and local connectivity in trafficking movements, as discussed in prior work~\cite{giommoni2017illicit}. Let $deg_{\hat{G}_i}$ denote the average degree after sparsification. For a fractional connectivity parameter $\tau_d^i \in (0,1]$, we require
\begin{equation*}
\deg(u) \ge \tau_d^i \cdot deg_{\hat{G}_i}.
\end{equation*}
Note that smaller values of $\tau_d^i$ allow sparser local structures, while larger values enforce denser local connectivity. Since the true local connectivity of the trafficking network is unobserved, we vary $\tau_d^i$ over a broad range and treat it as a parameter. For each node $u \in V$ that violates this condition, edges are added to nearby nodes until the threshold is satisfied. 
\end{enumerate}

\medskip

\noindent Let $E_i$ be the resulting edge set obtained for parameters $\tau_p^i$ and $\tau_d^i$.

\subsection{Augmenting trafficking networks with flow-capacities}
\label{sec:est_flow}

For each network graph generated in Section~\ref{sec:est_cap}, we estimate edge-level capacities so that the resulting flows are consistent with observed regional trafficking volumes. Our approach proceeds in two stages: (i) We construct lower bounds on edge flows to prevent degenerate solutions and ensure that all edges retain non-zero capacity. (ii) We compute a feasible flow assignment that satisfies regional volume constraints while respecting these lower bounds. This is achieved by solving the linear program shown in Figure~\ref{fig:network_flows}, which identifies edge capacities consistent with both structural and data-driven constraints.

\medskip

\noindent \textbf{\emph{Estimating edge-level minimum flows.}}
Intuitively, each edge is assigned a minimum flow based on the regional flow supported by its incident nodes, ensuring that all edges retain non-zero flow in the resulting network. Let $M_R = \left| {(u,v)\in E_i ;:; u\in R \ \text{or} \ v\in R } \right|$ denote the number of edges incident to region $R$. For each node $u \in R$, let $\deg(u)$ denote its degree. We define the node-level minimum-flow scale associated with node $u$ as
$$\textstyle \mu_u = \frac{\hat{g}(R)}{M_R} \cdot \deg(u),$$
which distributes the regional volume proportionally across incident edges. Next, we define a baseline lower bound for each edge $(u,v)\in E_i$ as
$$\textstyle m_{uv} = \alpha^i \min(\mu_u,\mu_v),$$
where $\alpha^i \in [0,1]$ is a scaling parameter that controls the overall magnitude of minimum flows. Small values of $\alpha^i$ allow the LP to allocate flows flexibly while still avoiding degenerate zero-capacity edges, whereas larger values impose stronger lower-bound requirements on edge usage. Since edge-level trafficking volumes are unobserved, we model $\alpha$ as a network-generation parameter and vary it over several orders of magnitude to capture uncertainty in how regional trafficking volume is distributed.

\smallskip

Finally, to induce variability across generated network flows, we introduce stochastic perturbations in these lower bounds. Specifically, the edge-level minimum flow is sampled as
$$
\ell_{uv} \sim 
\mathrm{Uniform}\!\left(
\max\{(1-\delta^i)m_{uv},0\}, \,
(1+\delta^i)m_{uv}
\right),
$$
where $\delta^i \in (0,1)$ controls the extent of perturbation and is treated as a parameter.

\begin{figure}[t!]
\begin{tcolorbox}[
    colback=gray!10,
    colframe=gray!50,
    boxrule=0.8pt,
    left=3pt, right=3pt, top=3pt, bottom=3pt
]
\textbf{Network Flow LP}
\vspace{-1em}
\[
\setlength{\jot}{2pt}
\begin{aligned}
\min \quad & f_{\mathrm{total}}(s) \\
\text{s.t.}\quad
& \forall v \in V,\ v \neq s,t:\ 
  \sum_{(u,v)\in E} f(u,v) = \sum_{(v,u)\in E} f(v,u) \\
& \forall R \in \mathcal{R}:\ 
  \sum_{v\in R} f_{\mathrm{total}}(v) \in [(1-\epsilon)g(R),(1+\epsilon)g(R)] \\
& \forall v \in V,\ v \neq t:\ 
  f(v) = \sum_{(v,w)\in E} f(v,w) \\
& \forall (u,v)\in E:\ f(u,v) \ge \ell_{u,v} \\
& \forall v\in V:\ f(v) \ge 0 \\
& \epsilon \ge 0
\end{aligned}
\]
\end{tcolorbox}
\vspace{-1em}
\caption{Linear programming (LP) for approximating edge flows from regional flows.}
\label{fig:network_flows}
\end{figure}

\medskip

\noindent \textbf{\emph{Estimating network edge-capacities satisfying regional constraints.}} We model trafficking activity as a flow from a source node $s$ to a sink node $t$ in the directed graph $G_i=(V,E_i)$. A flow function $f:E_i \rightarrow \mathbb{R}_{\ge 0}$ assigns non-negative values to edges and satisfies flow conservation at all intermediate nodes. For each node $v \in V \setminus \{s,t\}$, the total incoming flow equals the total outgoing flow. Let
$$ \textstyle f_{\text{total}}(v) = \sum_{(u,v)\in E_i} f(u,v)$$
denote the total incoming flow into node $v$.

Regional observations impose aggregate constraints on these node-level flows. Specifically, for each region $R \in \mathcal{R}$, the total incoming flow across nodes in $R$ should approximate the regional volume. To account for uncertainty in the data and prior imputation, we allow a multiplicative tolerance. A flow $f$ is said to be \emph{$\epsilon$-consistent} with the regional volumes if
$$ \frac{|g(R) - \sum_{v\in R} f_{\text{total}}(v)|}{g(R)} \leq \epsilon, \quad \forall R \in \mathcal{R}.
$$

Our objective is therefore to compute among all feasible solutions, the one that minimizes the total flow originating at the source. To ensure that every edge retains non-zero capacity, we impose lower bounds $\ell_{uv}$ on edge flows, as constructed above. The resulting problem can be formulated as a linear program (Figure~\ref{fig:network_flows}). We then determine the smallest value of $\epsilon$ for which the LP is feasible using binary search, repeatedly solving the network flow LP until the minimum feasible $\epsilon$ is identified. Let $\bar{f}_i(\cdot)$ denote the estimated trafficking capacities on the edges of the generated network.

\subsection{Data-Consistency based Filtering of the Ensemble of Networks}
\label{sec:generation_ablation}

The network generation procedure generates an ensemble of plausible trafficking networks under varying modeling parameters and stochastic realizations. However, not all generated networks are equally consistent with observed regional trafficking data. To ensure that downstream interdiction analysis is performed on realistic scenarios, we introduce a data-consistency based filtering step that selects a subset of networks whose induced flows closely match empirical observations.

\medskip

\noindent \textbf{Data-Consistency Metric.} We evaluate the quality of the induced regional flows in the ensemble of networks with the CCDB observations over the set of regions $\mathcal{C}_r$ with available data as
$$
\text{\emph{Average Relative Error, }ARE}(\bar{f}_i) = \frac{1}{|\mathcal{C}_r|} \sum_{R \in \mathcal{C}_r} \left| \frac{\hat g(R,\bar{f}_i) - g(R)}{g(R)} \right|, \quad where \quad \hat g(R,\bar{f}_i) = \sum_{v \in R} \sum_{(u,v)\in E_i} \bar{f}_i(u,v).
$$
Lower values of $\text{ARE}(\bar{f}_i)$ indicate stronger agreement between the synthesized network flows and empirical regional volumes.

\medskip

\noindent \textbf{Filtering Procedure.}
Given an ensemble of generated networks, we retain only those instances whose consistency error falls below a prescribed threshold. Formally, for a tolerance level $\eta$, we define the set of \emph{data-consistent networks} as
$$
\mathcal{G}^* = \{ (G_i,\bar{f}_i) : \text{ARE}(\bar f_i) \le \eta \}.
$$
This filtered ensemble represents the subset of plausible network realizations that are consistent with observed regional flow patterns, which serves as the scenario set for subsequent interdiction analysis in simulation-based evaluation of interdiction strategies under uncertainty.

\begin{figure}[t!]
\centering
\begin{minipage}{0.48\linewidth}
    \includegraphics[width=\linewidth]{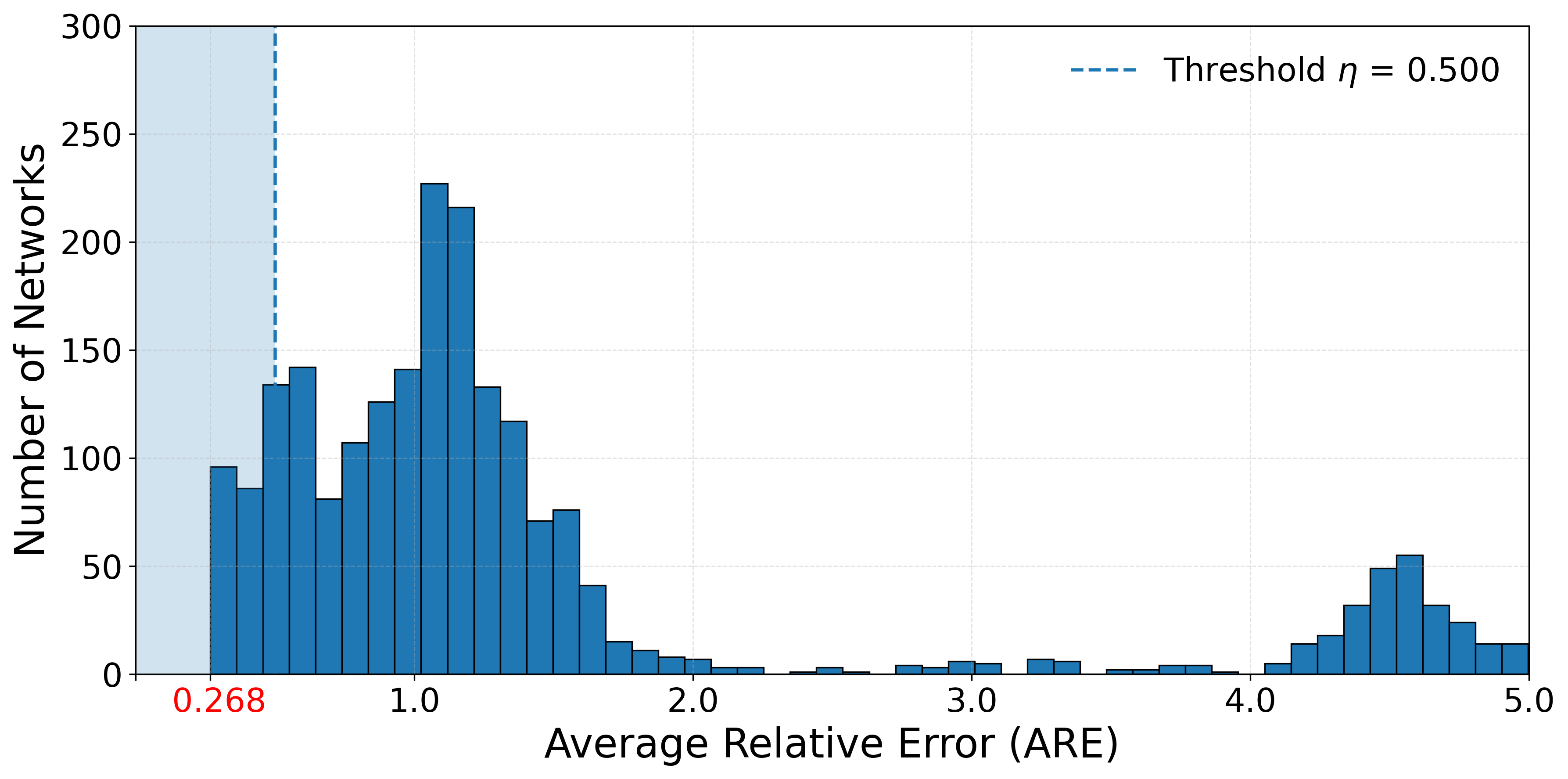}
\end{minipage}
\begin{minipage}{0.48\linewidth}
    \includegraphics[width=\linewidth]{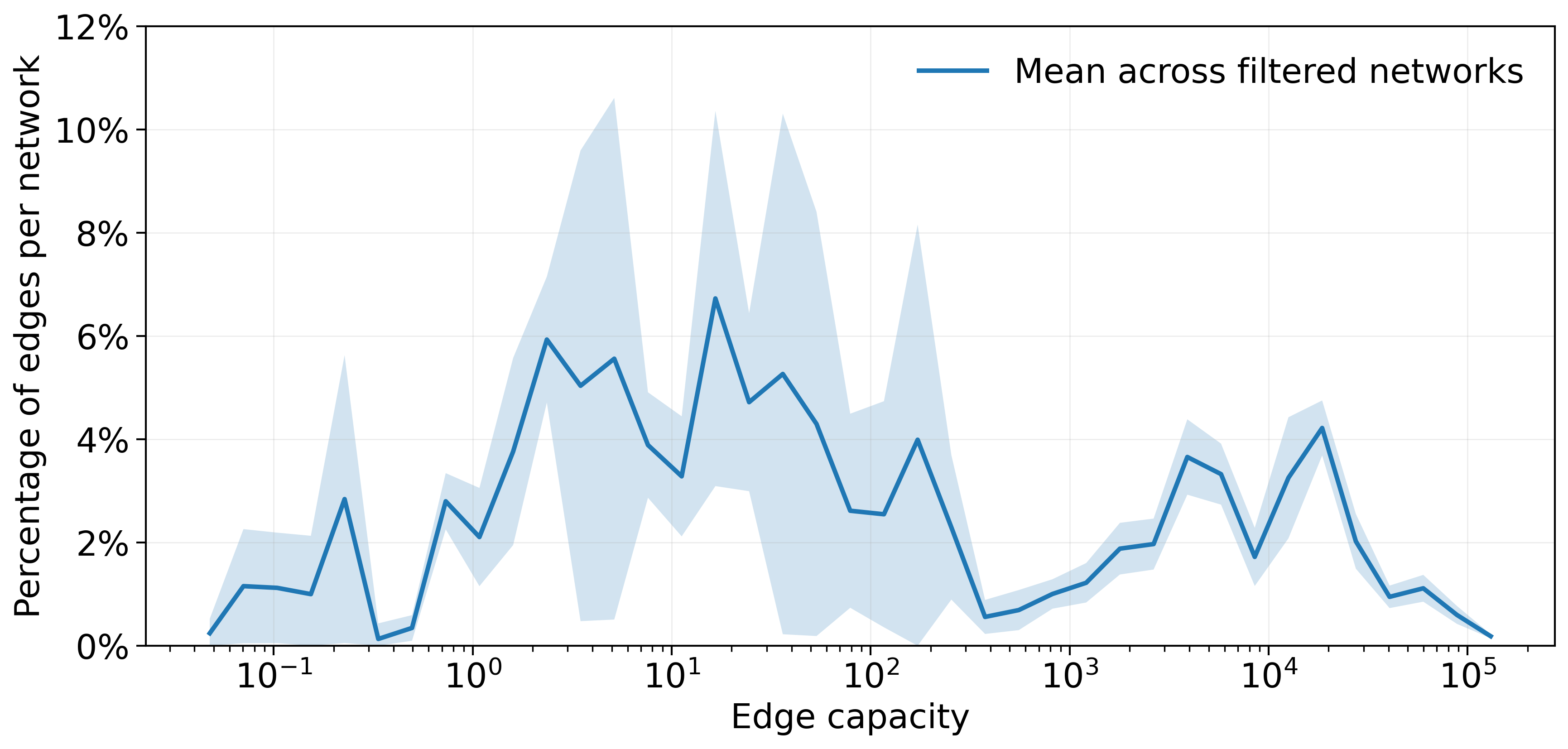}
\end{minipage}
\begin{minipage}{0.99\linewidth}
    \centering
    \includegraphics[width=0.24\linewidth]{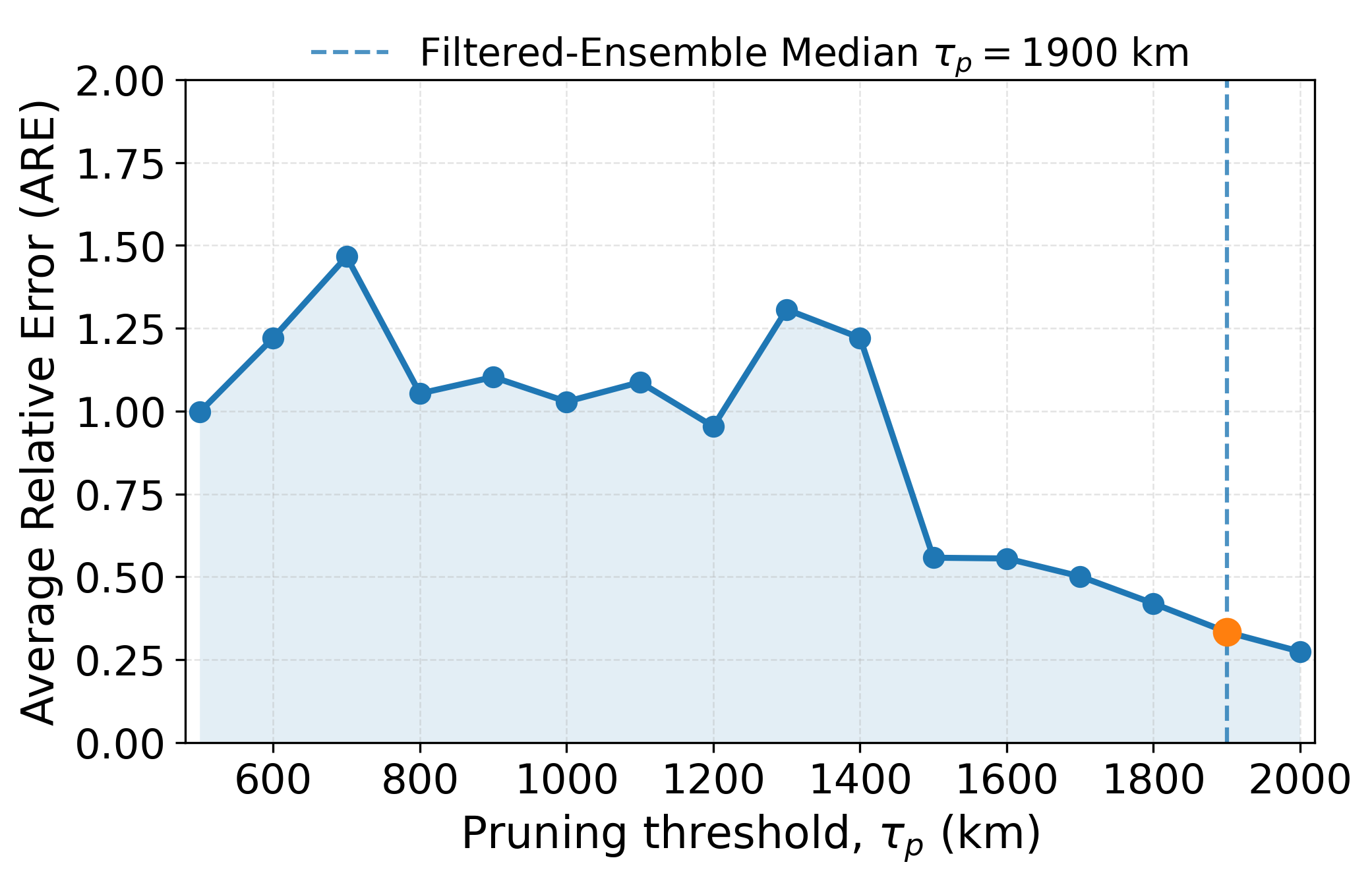}
    \includegraphics[width=0.24\linewidth]{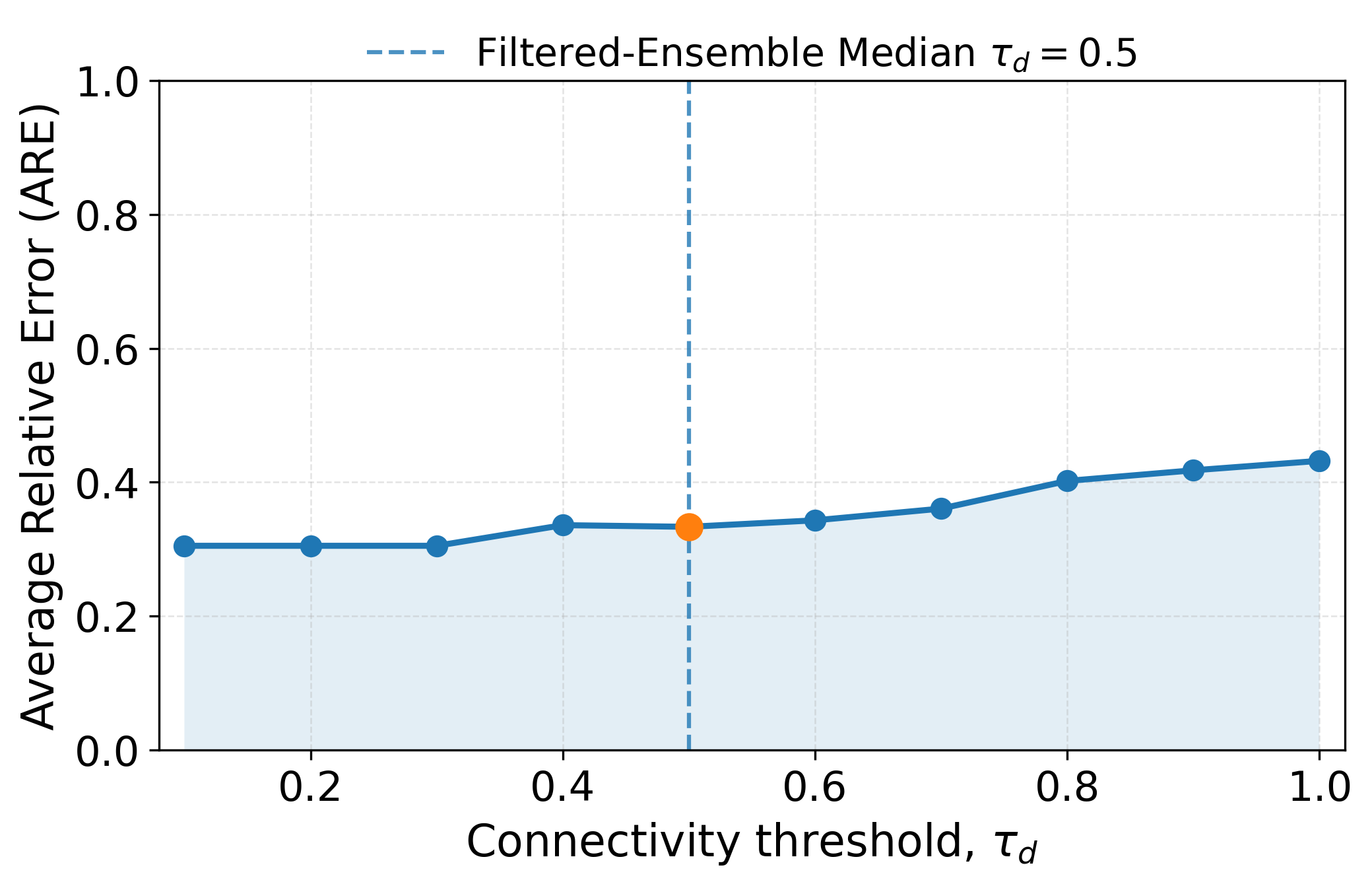}
    \includegraphics[width=0.24\linewidth]{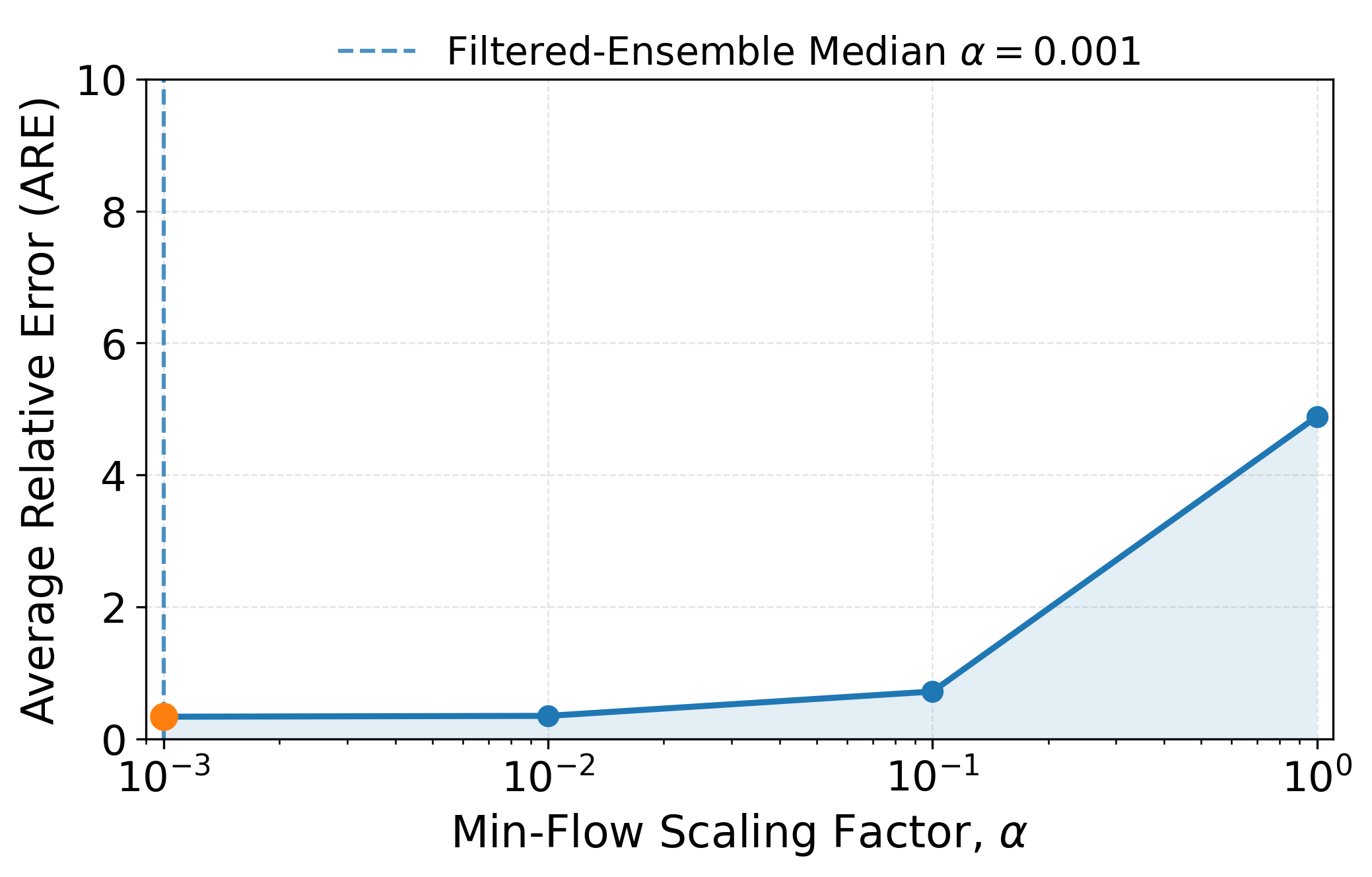}
    \includegraphics[width=0.24\linewidth]{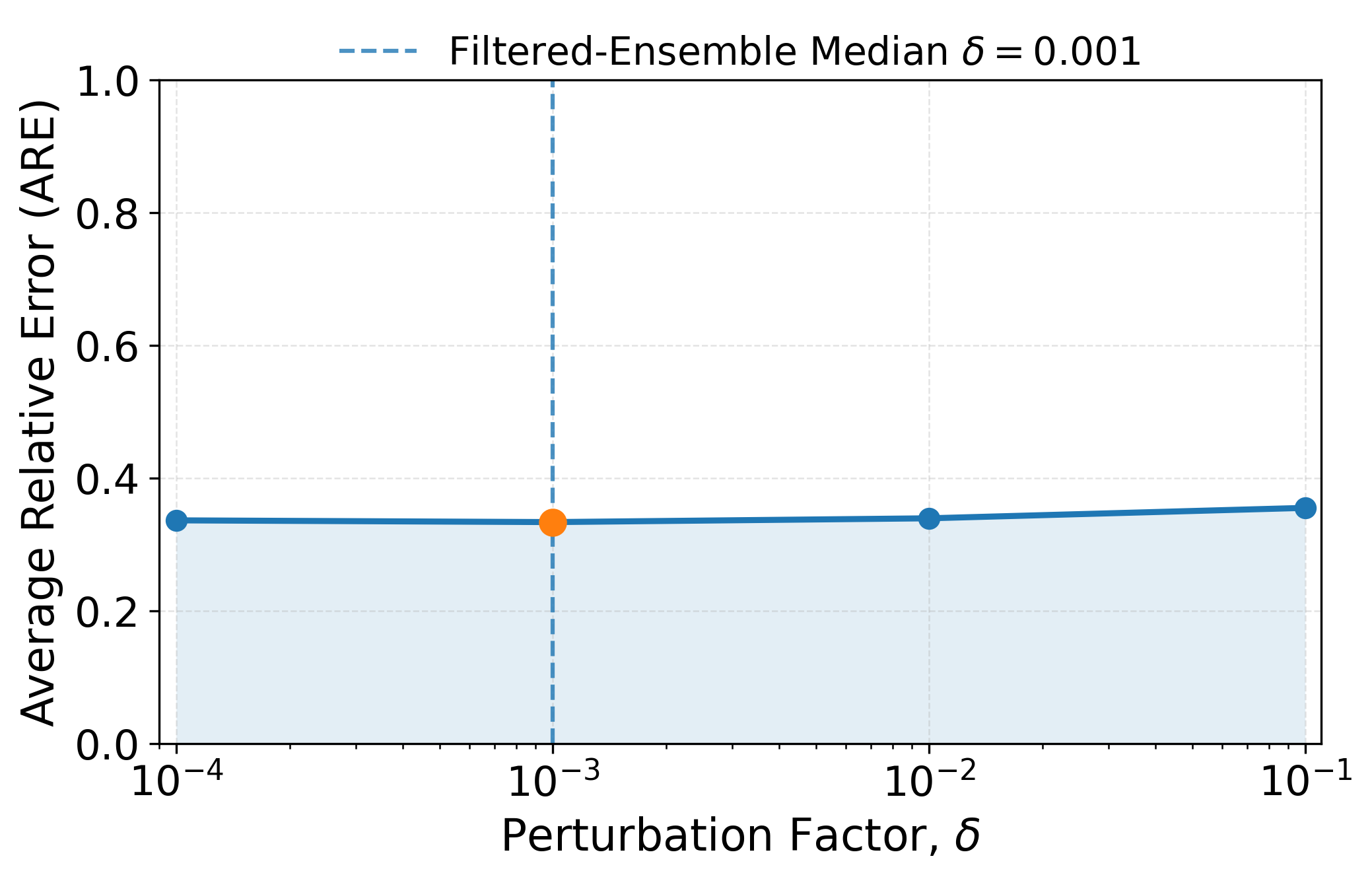}
\end{minipage}
\caption{(Top) Distribution of Average Relative Error (ARE) between reconstructed network flows regional aggregates and CCDB observations across the generated ensemble. The shaded region marks the Data-consistent networks (used for interdiction). (Bottom) Sensitivity of ARE to key network generation parameters.}
\label{fig:ablation_network_generation}
\end{figure}

We filter the ensemble of networks generated using simulation scenarios from Table~\ref{tab:network_generation_parameters} for ARE threshold $\eta = 0.50$, resulting in an ensemble of $230$ networks from the original ensemble of $2560$ networks. We illustrate the ARE distribution of the resulting ensemble of networks after and before filtering in Figure~\ref{fig:ablation_network_generation}.



\medskip

\noindent \textbf{Parameter Sensitivity Analysis.}
To understand how modeling parameters affect data consistency, we conduct a one-factor-at-a-time analysis over the parameter ranges specified in Table~\ref{tab:network_generation_parameters}. We fix the remaining network parameters to median value from set $\mathcal{G}^*$ at $\eta = 0.50$, which were $\tau_p = 1900$ km, $\tau_d = 0.5$ and $\alpha = \delta = 0.001$ and generate networks for this study. Figure~\ref{fig:ablation_network_generation} summarizes the sensitivity of the data-consistency metric to each parameter. This analysis provides insight into how the modeling parameters influence data-consistency of the generated networks.

\section{Solving Node Interdiction problems}\label{sec:node_interdiction}
The binary and robust flow interdiction problems defined in Section~\ref{sec:prelim} provide a stylized abstraction of real-world counter-narcotics strategies. While these formulations simplify operational complexities, they capture the essential tradeoffs between resource allocation and disruption of illicit flows, and thus provide a principled basis for systematic analysis. We formulate both problems as integer linear programs (ILPs). These formulations enable consistent simulation of interdiction strategies across the ensemble of synthesized networks.

\begin{figure}[b!]
\vspace{10pt}
\begin{tcolorbox}[
    colback=gray!10,
    colframe=gray!50,
    boxrule=0.8pt,
    left=4pt, right=4pt, top=4pt, bottom=4pt
]
\begin{minipage}[t]{0.48\columnwidth}
\centering
\textbf{\probbinarynode{}}
\[
\setlength{\jot}{2pt}
\begin{aligned}
\min \quad 
    & \sum_{e\in E} cap(e)\, y_e \\
\text{s.t.}\quad
    & \sum_{v\in V} c_v\, w_v \le B \\
    & y_e + z_e \ge d_u - d_v,\ \forall e \\
    & y_e + z_e \ge d_v - d_u,\ \forall e \\
    & d_s = 0,\ d_t = 1 \\
    & w_s = 0,\ w_t = 0 \\
    & w_u + w_v \ge z_e,\ \forall e \\
    & w_v, d_v \in \{0,1\},\ \forall v \\
    & y_e, z_e \in \{0,1\},\ \forall e
\end{aligned}
\]
\end{minipage}
\hfill
\begin{minipage}[t]{0.48\columnwidth}
\centering
\textbf{\probrobustbinarynode{}}
\vspace{-0.3em}
\[
\setlength{\jot}{2pt}
\begin{aligned}
\min \quad & \alpha \\
\text{s.t.}\quad
& \sum\limits_{e\in E_i} cap(e)\, y^i_e \le \alpha \\
& \sum_{v\in V} c_v\, w_v \le B \\
& y^i_e + z^i_e \ge d^i_u - d^i_v,\ \forall e,i \\
& y^i_e + z^i_e \ge d^i_v - d^i_u,\ \forall e,i \\
& d^i_s = 0,\ d^i_t = 1,\ \forall i \\
& w_s = 0,\ w_t = 0 \\
& w_u + w_v \ge z^i_e,\ \forall e,i \\
& w_v, d_v \in \{0,1\},\ \forall v \\
& y^i_e, z^i_e \in \{0,1\},\ \forall e,i
\end{aligned}
\]
\end{minipage}
\end{tcolorbox}
\vspace{-1em}
\caption{ILP formulations for the \probbinarynode{} and \probrobustbinarynode{} problems.}
\label{fig:LP_node_cost}

\end{figure}

\medskip

\noindent We propose an ILP solution to the\textbf{~\probbinarynode{}} problem in Figure~\ref{fig:LP_node_cost}, which uses the following decision variables:
\begin{itemize}
\item 
$w_v\in\{0, 1\}$ for $v\in V$, with $w_v=1$ if node $v$ is interdicted
\item 
$z_e \in \{0,1\}$ for each $e\in E$: $z_e=1$ if an end point of $e$ is interdicted
\item $d_v \in \{0,1\}$ for each $v\in V$: cut indicator (0 = $s$-side, 1 = $t$-side).
\item $y_e  \in \{0,1\}$ for each $e=(u, v)\in E$: $y_e=1$ if the end points of $e$ are on different sides of the cut, i.e., $d(u)\neq d(v)$, and neither end point is interdicted.
\end{itemize}

\smallskip

\begin{observation}
The ILPs in Figure \ref{fig:LP_node_cost} give optimal solutions to the  \probbinarynode{} and \probrobustbinarynode{} problems.
\end{observation}


\noindent For the \textbf{\probrobustbinarynode{}} problem, we consider cut-LP constraints for each graph $G_i$ (as in Figure \ref{fig:LP_node_cost}), but the nodes chosen for interdiction are common.
Figure \ref{fig:LP_node_cost} shows the ILP for this problem.



\subsection{Empirical Study on Ensemble of Networks}

We now evaluate interdiction strategies using the ensemble of synthesized networks, with the goal of understanding how network uncertainty influences both solution quality and stability. In particular, we use the ILP formulations above as decision models within a simulation-based evaluation framework, where each network realization represents a plausible trafficking scenario.

For the filtered ensemble of network synthesized in Section~\ref{sec:framework}, we solve the corresponding ILP to obtain optimal as well as robust interdiction decisions and evaluate their impact on residual flow. We assume uniform node interdiction costs due to the absence of reliable cost data. 

\medskip


\begin{wrapfigure}{r}{0.45\textwidth}
    \vspace{-0.2em}
    \centering
    \includegraphics[width=0.45\textwidth]{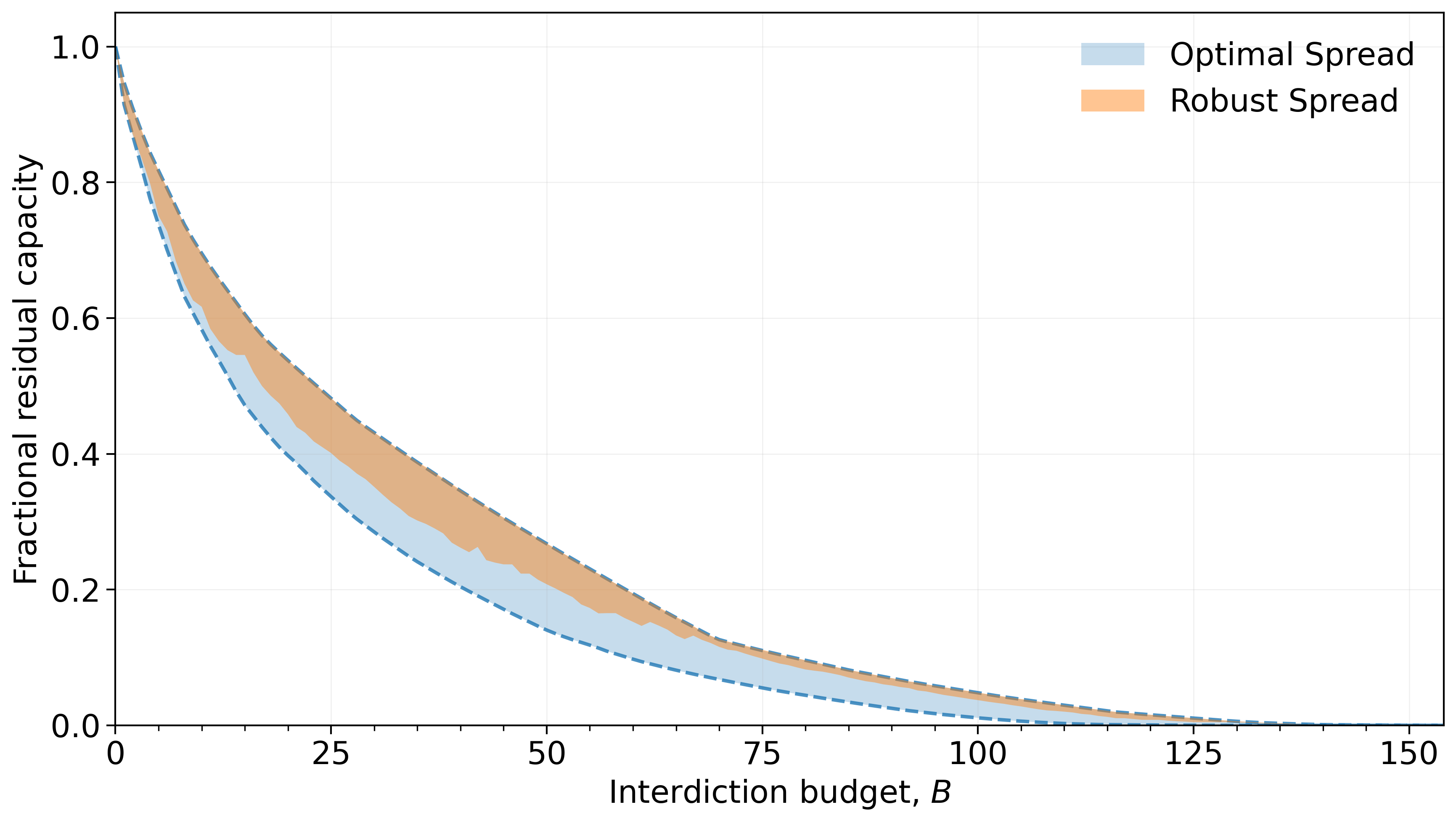}
    \caption{Residual Capacity across interdiction decisions.}
    \label{fig:robust_fractional_residual}
    \vspace{-2em}
\end{wrapfigure}

\subsubsection{Interdiction yields rapid initial reductions in Network Capacity}

Figure \ref{fig:robust_fractional_residual} shows the fractional residual flow (normalized by the initial maximum flow from $s$ to $t$) vs the interdiction budget (total number of nodes in the graph). We illustrate the maximum and minimum residual capacity at per network decision across the ensemble, which we refer to as \textbf{spread}, capturing the sensitivity of interdiction outcomes to network scenarios. The optimal spread captures the variation in residual flows of the network-specific optimal interdiction decision. We observe that the optimal residual flows have a steep initial drop, which slows down gradually. This suggests the presence of critical bottleneck nodes whose removal disproportionately disrupts flow.

However, the spread in residual capacities across networks indicates that the effectiveness of these early interventions varies significantly across the ensemble of networks. This motivates a deeper investigation into the stability of interdiction decisions.

\subsubsection{Optimal interdiction is unstable under Network Uncertainty}

\begin{figure}[t!]
    \centering
    \includegraphics[width=0.98\linewidth, clip, trim= 0 10cm 0 8cm]{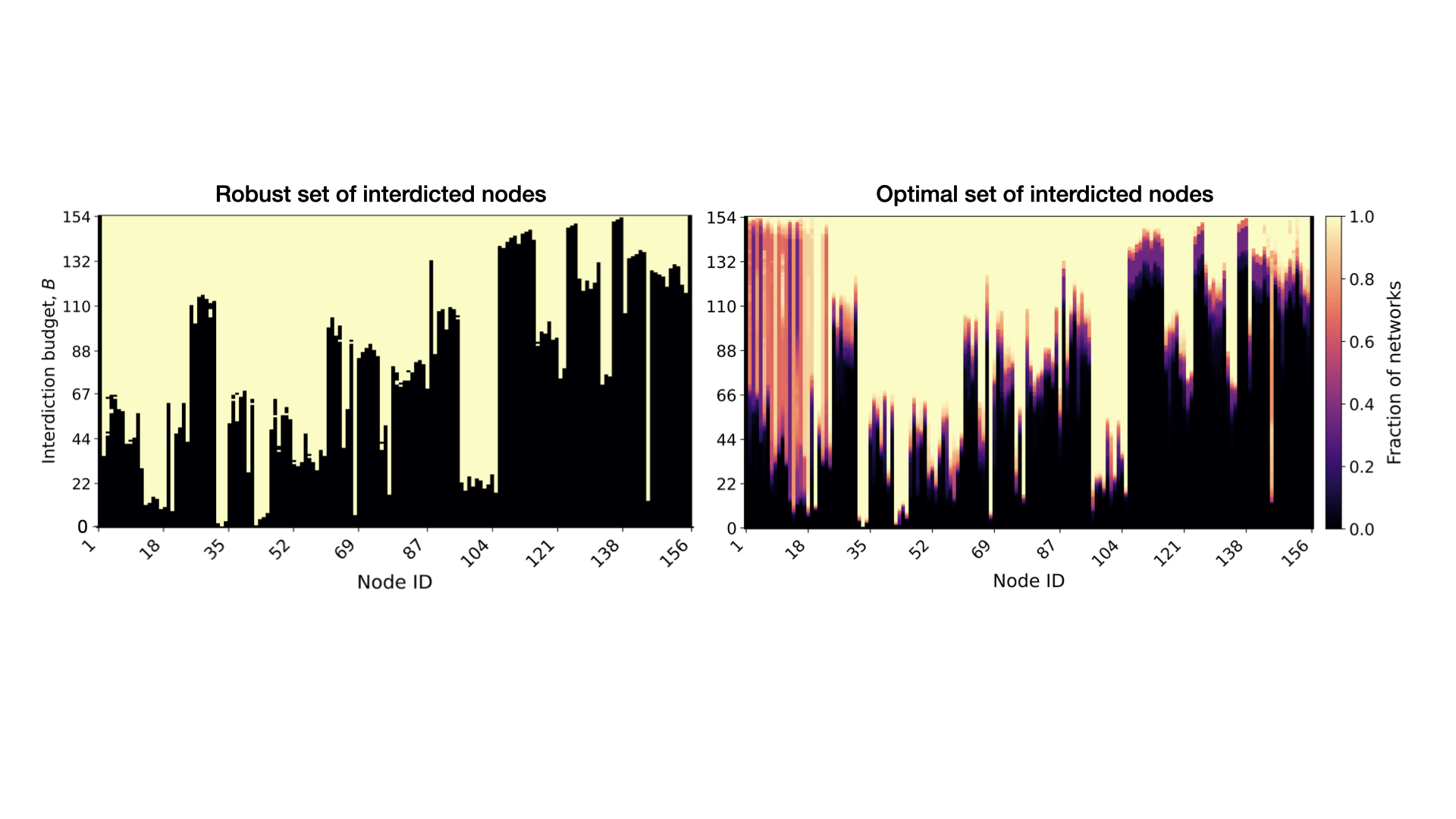}

    \caption{\textbf{Distribution of interdicted nodes over the filtered ensemble of networks.} (Left) The robust interdiction set is computed across all networks, resulting in binary selection (0 or 1) for each node. (Right) The optimal interdiction set is computed separately for each network, and we report the fraction of networks in which a node is selected at a given budget $B$.}
    \label{fig:interdiction}
\end{figure}

Figure~\ref{fig:interdiction} (Right) illustrates the distribution of interdicted nodes across the ensemble of network scenarios. The results show that mid-range interdiction nodes selected optimally for each network vary substantially, indicating that optimal interdiction strategies depend strongly on the underlying network characteristics. We include detailed geographical visualization of the interdiction decision in the Supplementary Information.

\begin{figure}[b!]
    \centering
    \includegraphics[width=0.95\linewidth, trim = 2cm 10.5cm 2cm 10cm, clip]{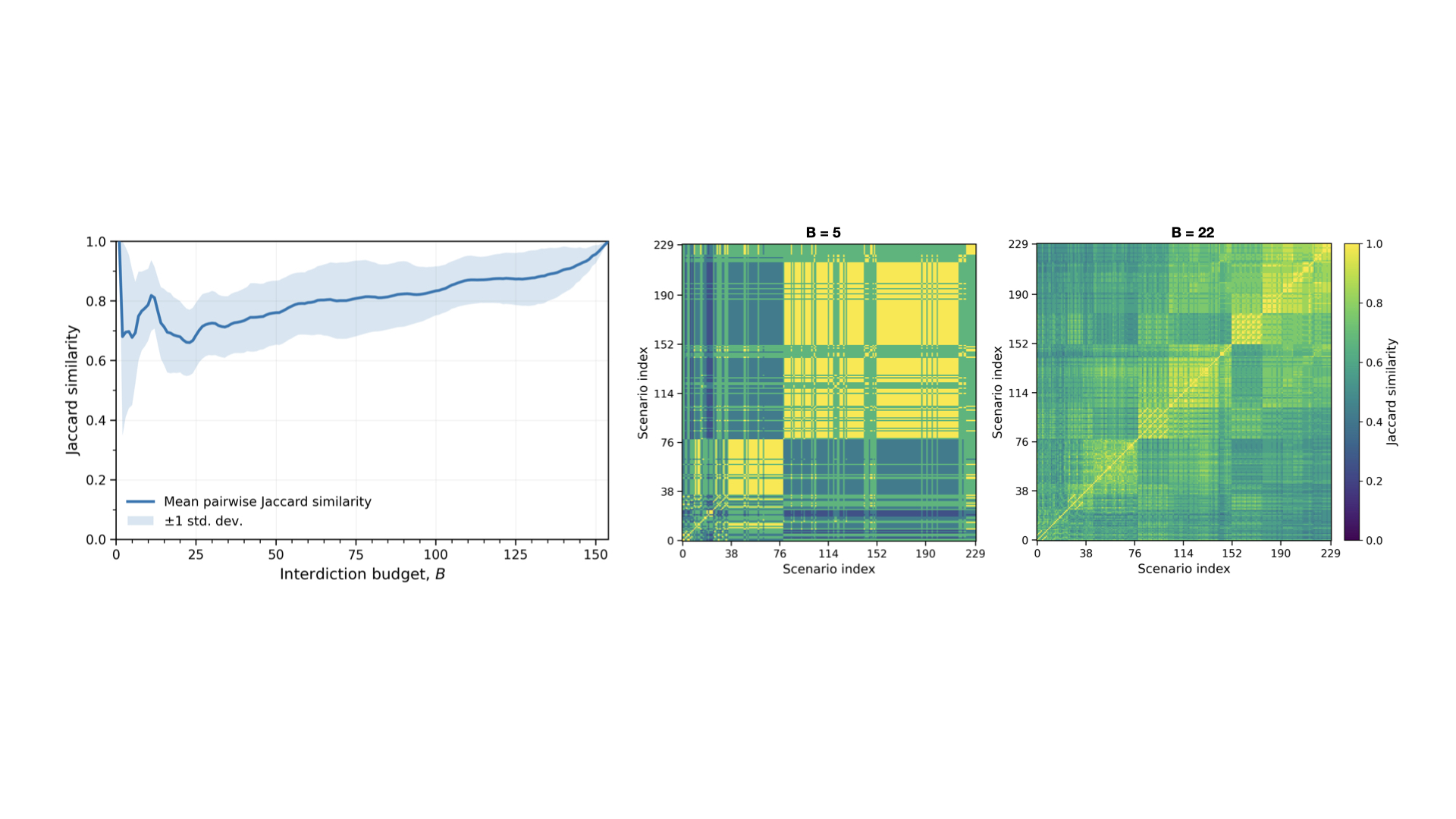}
    \caption{Jaccard Similarity for the optimal set of interdicted nodes. }
    \label{fig:NodeBudget_jaccard}
\end{figure}

We quantify the stability of interdiction decisions across the ensemble by comparing the sets of interdicted nodes selected at different budgets. Figure~\ref{fig:NodeBudget_jaccard} reports the mean pairwise similarity between interdicted node sets using the Jaccard Similarity, defined for any two scenarios $i$ and $j$ with interdicted node sets $S_i$ and $S_j$ as 
\[
J(S_i, S_j) = \frac{|S_i \cap S_j|}{|S_i \cup S_j|}.
\]
We observe that similarity is relatively low at small budgets, indicating that optimal decisions are highly sensitive to network scenarios in the low-resource regime. As the budget increases, similarity rises steadily, suggesting that larger budgets lead to more consistent node selection across scenarios. This trend reveals a fundamental tradeoff: while early interventions achieve high impact, they are also highly unstable, whereas higher budgets uncover a more stable set of structurally critical nodes that are consistently selected across networks.

\begin{figure}[t!]
    \centering
    \includegraphics[width=0.99 \linewidth, trim = 0 12.5cm 0 12cm, clip]{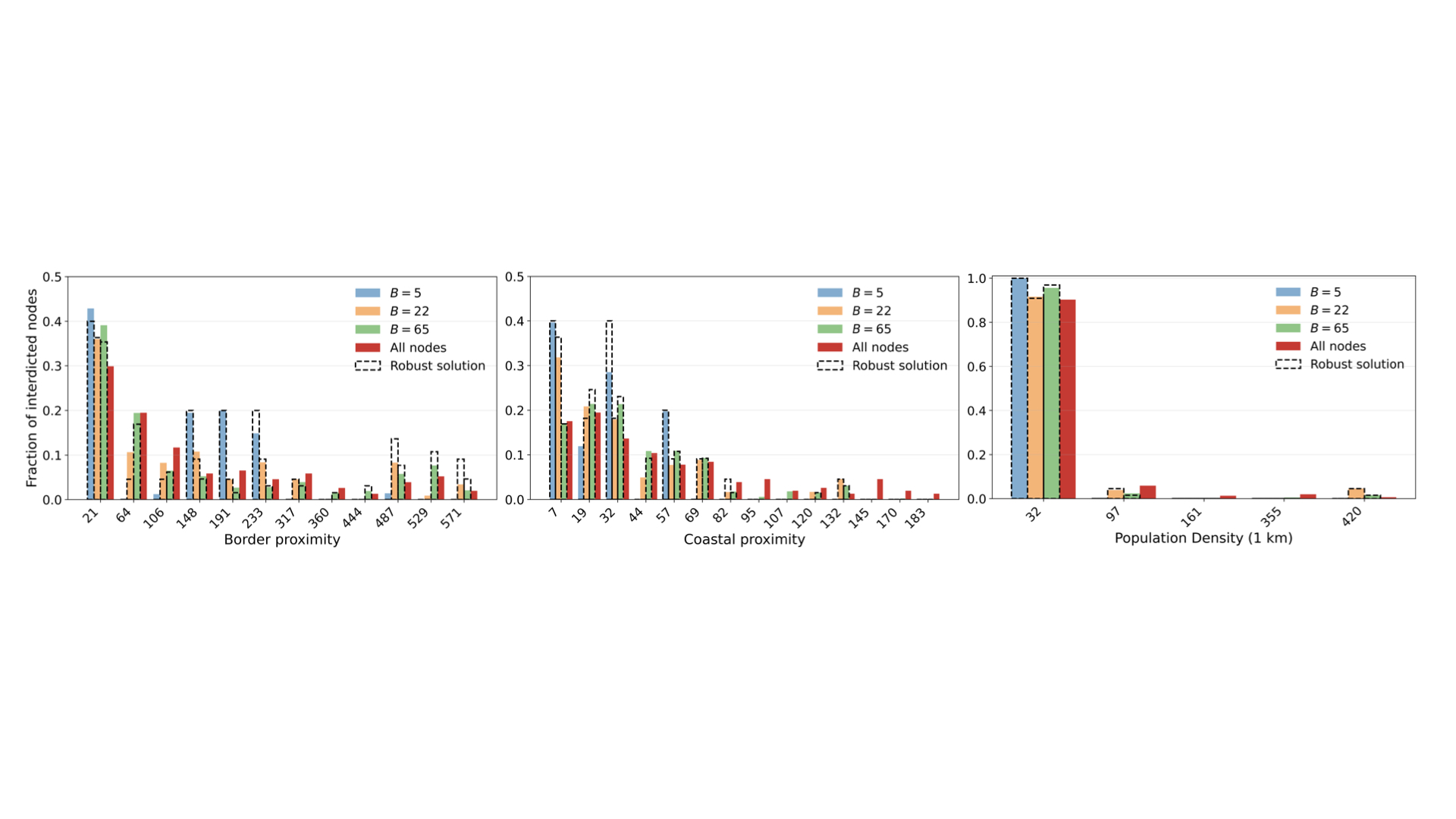}
    \caption{\textbf{Feature Distribution of interdicted nodes.} Comparison of features of all nodes with the mean set of optimal across the ensemble of network as well as robust interdicted nodes.
    }
    \label{fig:NodeBudget_feature}
\end{figure}

To characterize the properties of nodes selected in optimal interdiction solutions, we analyze the distributions of key geographic and demographic features associated with interdicted locations. Figure~\ref{fig:NodeBudget_feature} compares the feature distributions of interdicted nodes across multiple budgets with the baseline distribution over all nodes in the network. We observe that the distributions of interdicted nodes differ significantly from that of all nodes particularly for border proximity and coastal proximity, indicating that optimal interdiction strategies preferentially target geographically structured regions rather than uniformly distributed nodes. Moreover, the variation across budgets highlights that the characteristics of selected nodes evolve with resource availability, yet consistently remain distinct from the baseline distribution. This suggests that nodes selected in optimal solutions are structurally and functionally different from typical nodes in the network.

\subsubsection{Robust interdiction performs near-optimal, identifying critical nodes}

The variability of optimal solutions across network scenarios implies that a strategy that is optimal for one instance may perform poorly for another, posing a fundamental challenge under network uncertainty. This motivates the \probrobustbinarynode{} problem (Section~\ref{sec:node_interdiction}). Figure~\ref{fig:robust_fractional_residual} shows the fractional residual flow (normalized by the initial maximum flow from $s$ to $t$) as a function of the interdiction budget for robust interdiction decisions across the ensemble of networks.

We observe that, at lower budgets, the residual capacity achieved by the robust solution lies within the optimal spread, indicating near-optimal performance while maintaining stability across scenarios. At higher budgets, the robust solution becomes sub-optimal for individual network instances, reflecting the tradeoff between optimality and robustness. Furthermore, the characteristics of nodes selected in robust solutions differ from those of typical nodes in the network and vary with budget, consistent with the patterns observed for optimal solutions (Figure~\ref{fig:NodeBudget_feature}).

Despite network uncertainty, we observe the emergence of a core set of nodes that are consistently selected across budgets (Figure~\ref{fig:interdiction}). These nodes appear as persistent vertical bands in the optimal selection heatmap, indicating high selection frequency across network realizations. We define this core set as nodes that are selected with high frequency across various budgets. Their stability suggests that they correspond to structurally critical regions that act as bottlenecks in the trafficking network and remain invariant to perturbations in network topology and capacities.

\section{Concluding remarks}\label{sec:conclusions}
We studied robust network flow interdiction problems using realistic data pertaining to narcotics networks first reported in~\cite{magliocca2019modeling,magliocca2021comparative}. 
As a first step, we created an ensemble of narco-traffic networks building on the work of~\cite{magliocca2019modeling,magliocca2021comparative}. We then formulated and solved a robust node interdiction problem on the ensemble of the networks --- the goal was to try and account for the uncertainty resulting from significant data sparsity for constructing such networks. Computational experiments yield natural strategies for interdiction. 

While the present work focuses on establishing the robust interdiction framework and demonstrating its behavior on data-consistent network ensembles, future work will include a broader benchmark study against heuristic and single-network baselines, as well as a systematic sensitivity analysis of final interdiction decisions. Several other extensions can also be undertaken. These include: ($i$) studying the problem as a repeated game and designing simulation-assisted methods for solving such games; ($ii$) studying variants that account for the discrete and time-varying nature of trafficking activity; and ($iii$) developing new generative AI methods for synthesizing narcotic trafficking networks using diverse data sources.

\smallskip

\noindent \textbf{Acknowledgments:} We thank members of BI and NSDPI for a number of useful comments and discussions.  This research is based upon work partially supported in by the Office of the Director of National Intelligence (ODNI) via Contract No. 2024-24070100001 and the UVA Contagion Science program.

\clearpage

\appendix
\section*{Appendix}

\section{Approximating edge capacities using CCDB Dataset: Details \& Results}

\noindent \textbf{Estimating regional volumes from CCDB dataset.} The dataset provides drug-trafficking interdiction volumes over administrative regions in Central-American countries. The data is largely coarse spanning multiple year for some regions and missing for others. As a first step, we simplify the data to be the maximum drug-trafficking observed in each region across all recorded timesteps values. 

\begin{table}[h!]
\centering
\footnotesize
\begin{tabular*}{\textwidth}{@{\extracolsep{\fill}}lr@{\hspace{1cm}}lr}
\toprule
\textbf{Region} & \textbf{Volume} & \textbf{Region} & \textbf{Volume} \\
\midrule
\textbf{Guanacaste, CR} & 16525 & El Progreso, GT & 4200 \\
\textbf{Limón, CR} & 89827 & Retalhuleu, GT & 37900 \\
\textbf{Puntarenas, CR} & 167982 & Comayagua, HN & 200 \\
\textbf{Alta Verapaz, GT} & 4200 & Copán, HN & 20750 \\
\textbf{Izabal, GT} & 20750 & El Paraíso, HN & 1459 \\
\textbf{Petén, GT} & 16972 & La Paz, HN & 200 \\
\textbf{San Marcos, GT} & 37900 & Yoro, HN & 10320.75 \\
\textbf{Atlántida, HN} & 13740 & Boaco, NI & 34630 \\
\textbf{Choluteca, HN} & 1400 & Chinandega, NI & 1400 \\
\textbf{Colón, HN} & 25825 & Chontales, NI & 34630 \\
\textbf{Cortés, HN} & 200 & Estelí, NI & 13089 \\
\textbf{Gracias a Dios, HN} & 211954 & Jinotega, NI & 24778 \\
\textbf{Olancho, HN} & 1518 & Rivas, NI & 16525 \\
\textbf{Atlántico Norte, NI} & 48038 & Río San Juan, NI & 62228.5 \\
\textbf{Atlántico Sur, NI} & 34630 & Bocas del Toro, PA & 97562.33 \\
\textbf{Chiriquí, PA} & 34878 & Coclé, PA & 72935.5 \\
\textbf{Colón, PA} & 23570 & Emberá, PA & 155137 \\
\textbf{Darién, PA} & 155137 & Kuna Yala, PA & 85598.33 \\
\textbf{Panamá, PA} & 78088 & Ngöbe Buglé, PA & 78589.5 \\
\textbf{Veraguas, PA} & 122301 & Panamá Oeste, PA & 50829 \\
\textbf{Usulután, SV} & 1525 & Puntarenas, PA & 86784.15 \\
Alajuela, CR & 92253.5 & Santa Ana, SV & 1525 \\
Heredia, CR & 89827 &  &  \\
\bottomrule
\end{tabular*}

\caption{\small Drug trafficking volumes for ADM1 regions in the Narcologic network. Region names are shown as \emph{State, Country Code}. Volumes for regions in bold are provided by the CCDB dataset, while the remaining volumes are estimated using the neighboring-region procedure described above.}
\label{tab:adm1_volume_estimates}
\end{table}

\begin{figure}[h!]
    \centering
    \begin{subfigure}[b]{0.29\linewidth}
        \centering
        \includegraphics[width=\linewidth, trim = 0 0 4cm 0.8cm, clip]{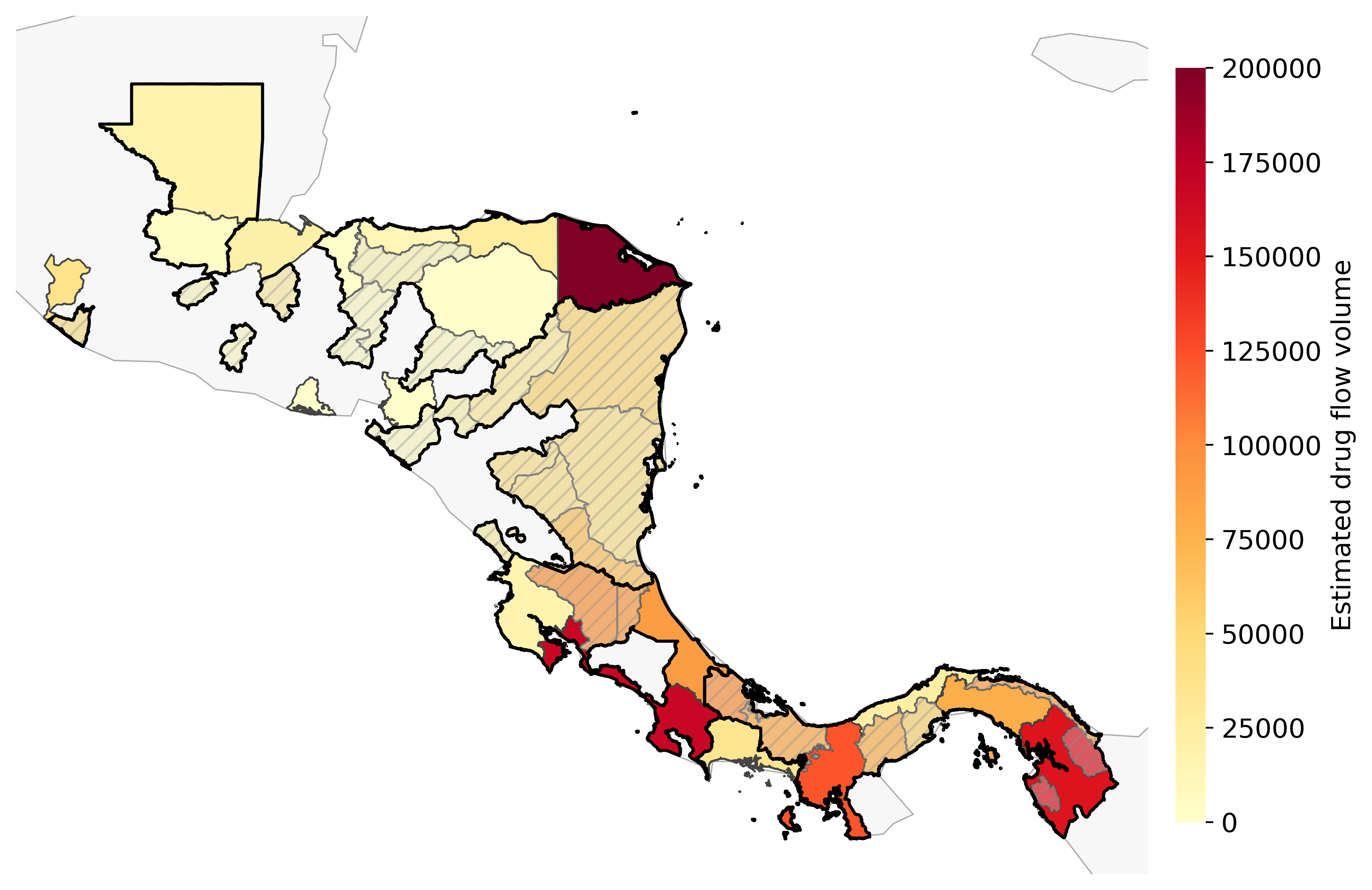}
        \caption{}
    \end{subfigure}
    \begin{subfigure}[b]{0.35\linewidth}
        \centering
        \includegraphics[width=\linewidth, trim = 0 0 0cm 0.8cm, clip]{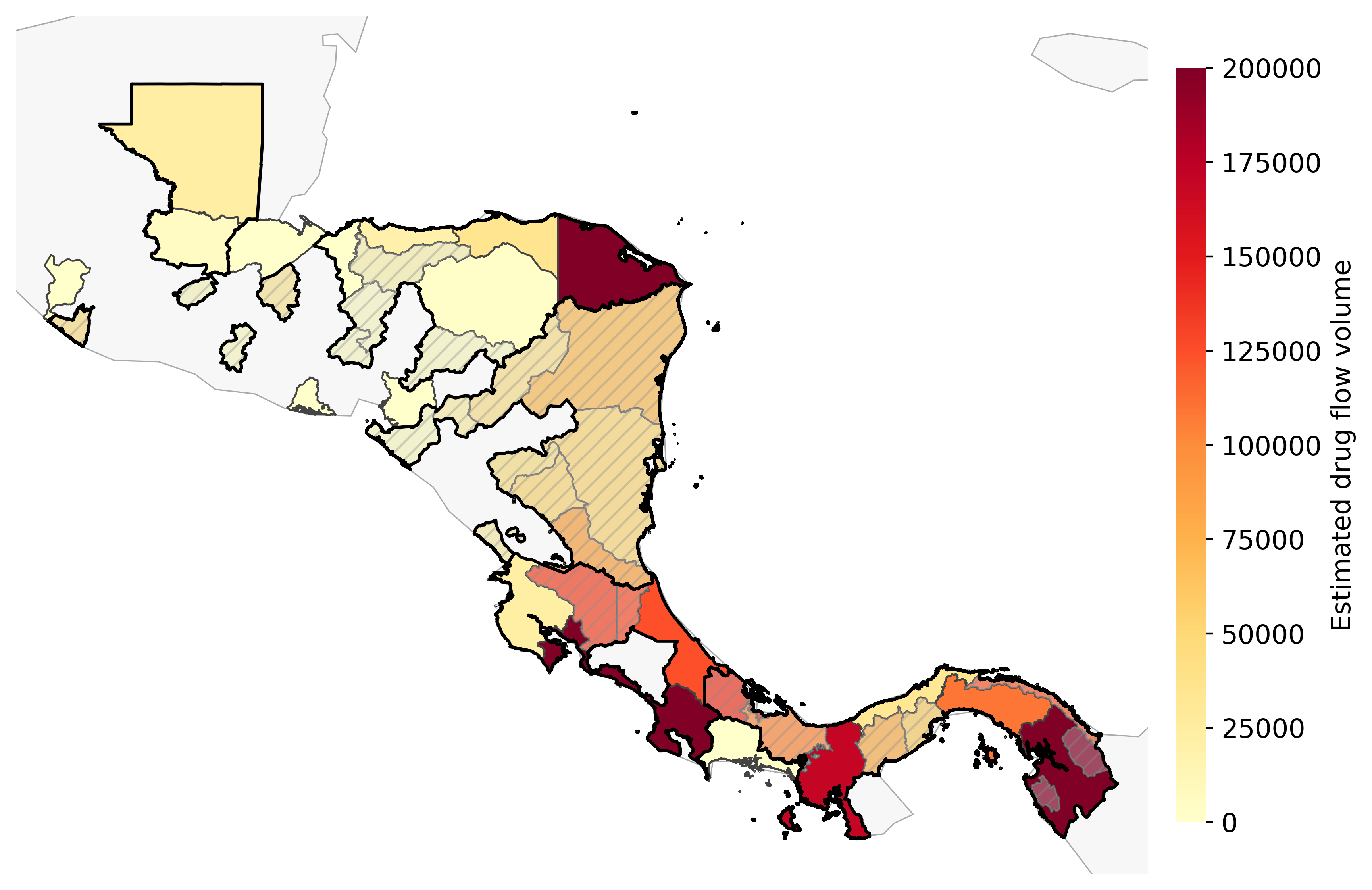}
        \caption{}
    \end{subfigure}
    \caption{\small\textbf{Observed and reconstructed regional trafficking capacities.} (a) Regional trafficking volumes from the CCDB dataset. (b) Regional inflows reconstructed from synthesized edge flows under the \emph{Baseline} network configuration. Hatched regions indicate administrative units without CCDB observations, where volumes in (a) were estimated using neighboring-region interpolation.}
    \label{fig:flow_region_approx}
\end{figure}

\clearpage
\section{Additional Empirical Details}

\noindent \textbf{Experimental setup.} All robust interdiction instances were formulated as mixed-integer linear programs and solved using the Gurobi Optimizer through its Python interface. Unless otherwise stated, we used Gurobi’s default automatic settings for core MIP procedures. The relative optimality gap tolerance was set to $10^{-4}$. 

\begin{figure}[h!]
    \centering
    \includegraphics[width=0.4\linewidth]{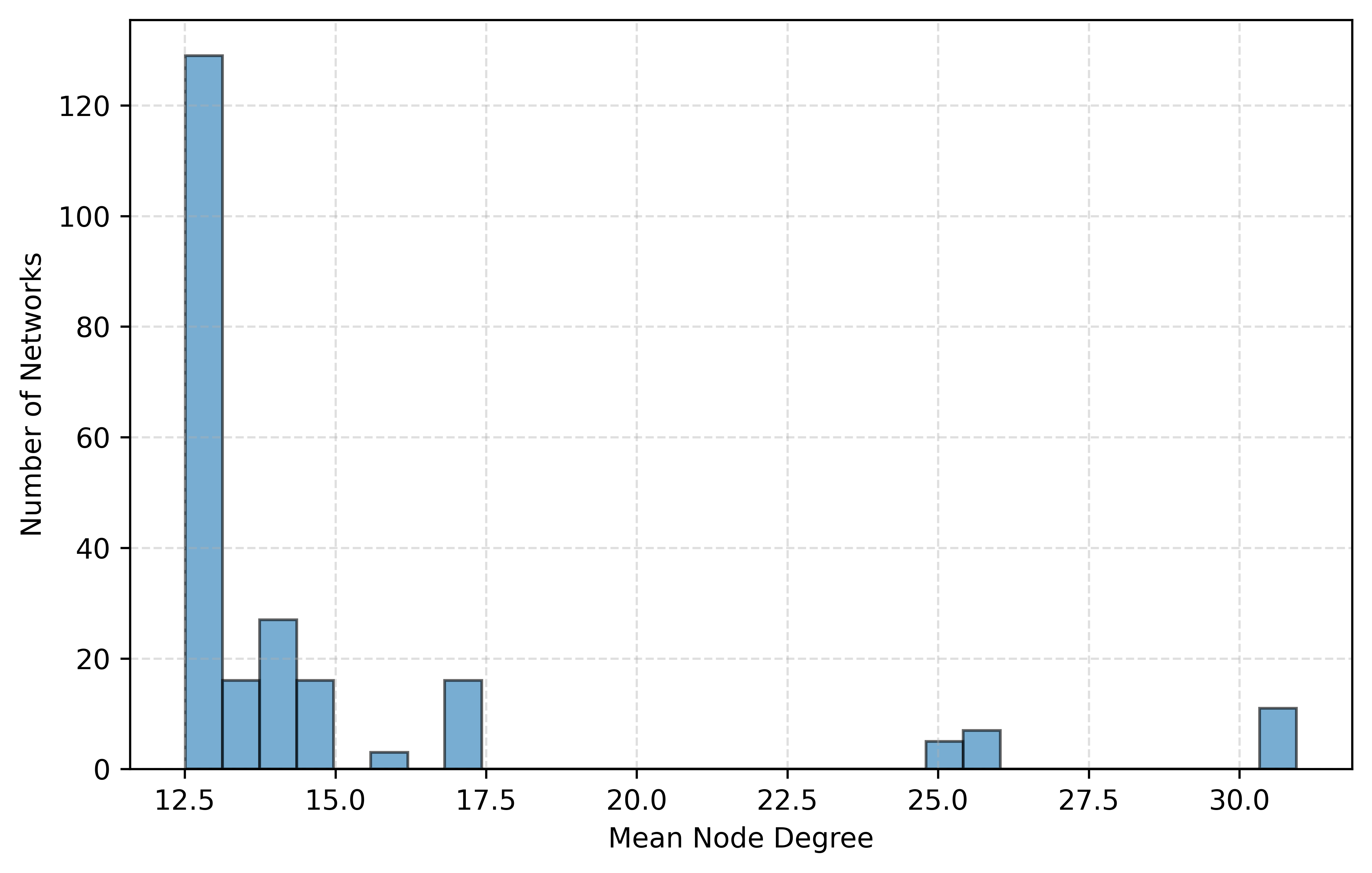}
    \includegraphics[width=0.4\linewidth]{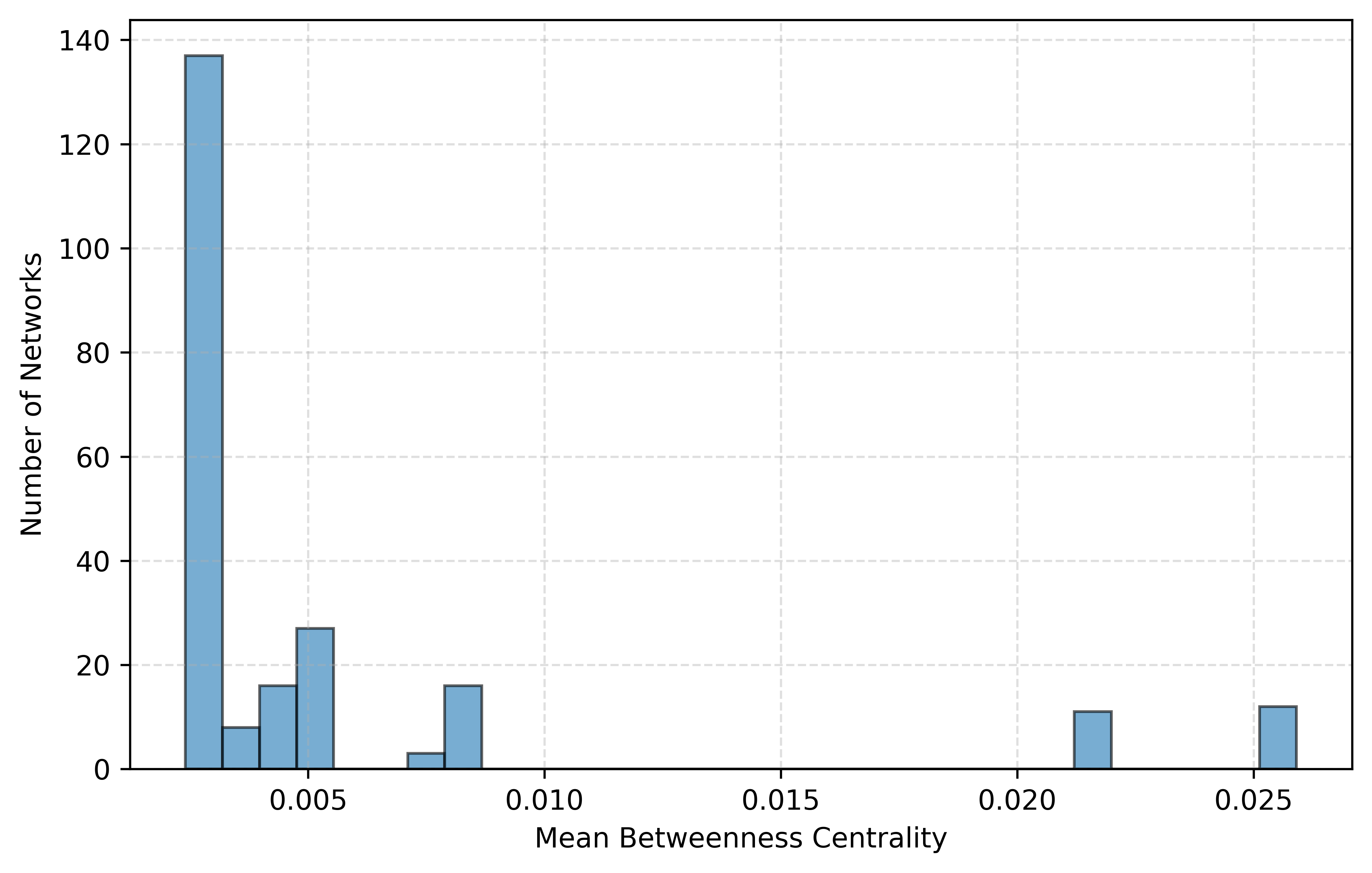}
    \caption{Distribution of network properties in the filtered ensemble of networks at $\eta = 0.50$.}
    \label{fig:dist_ensemble_props}
\end{figure}

\begin{figure}[b!]
    \centering

    \includegraphics[width=0.42\linewidth, clip, trim= 0 2cm 4cm 0]{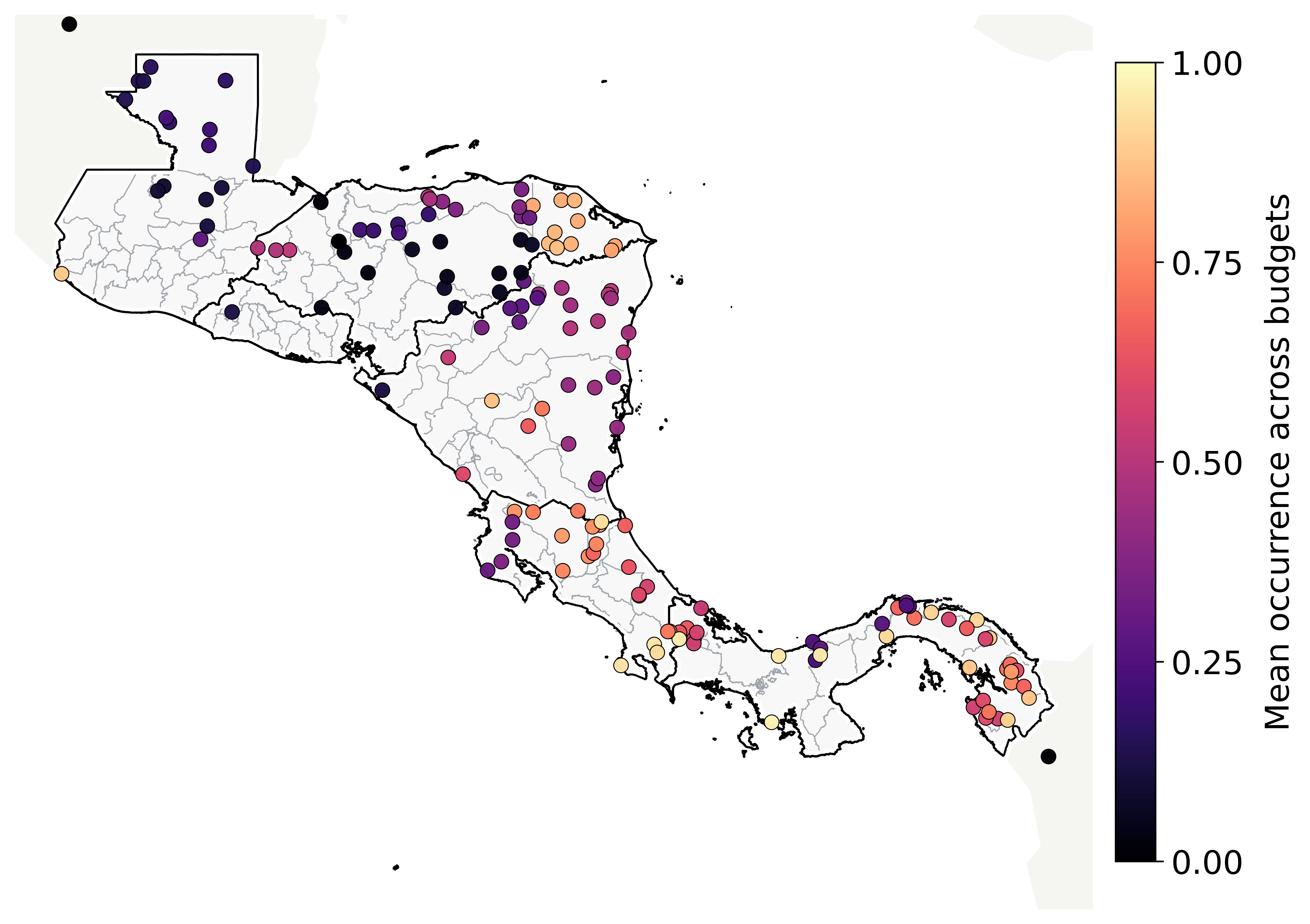}
    \includegraphics[width=0.49\linewidth]{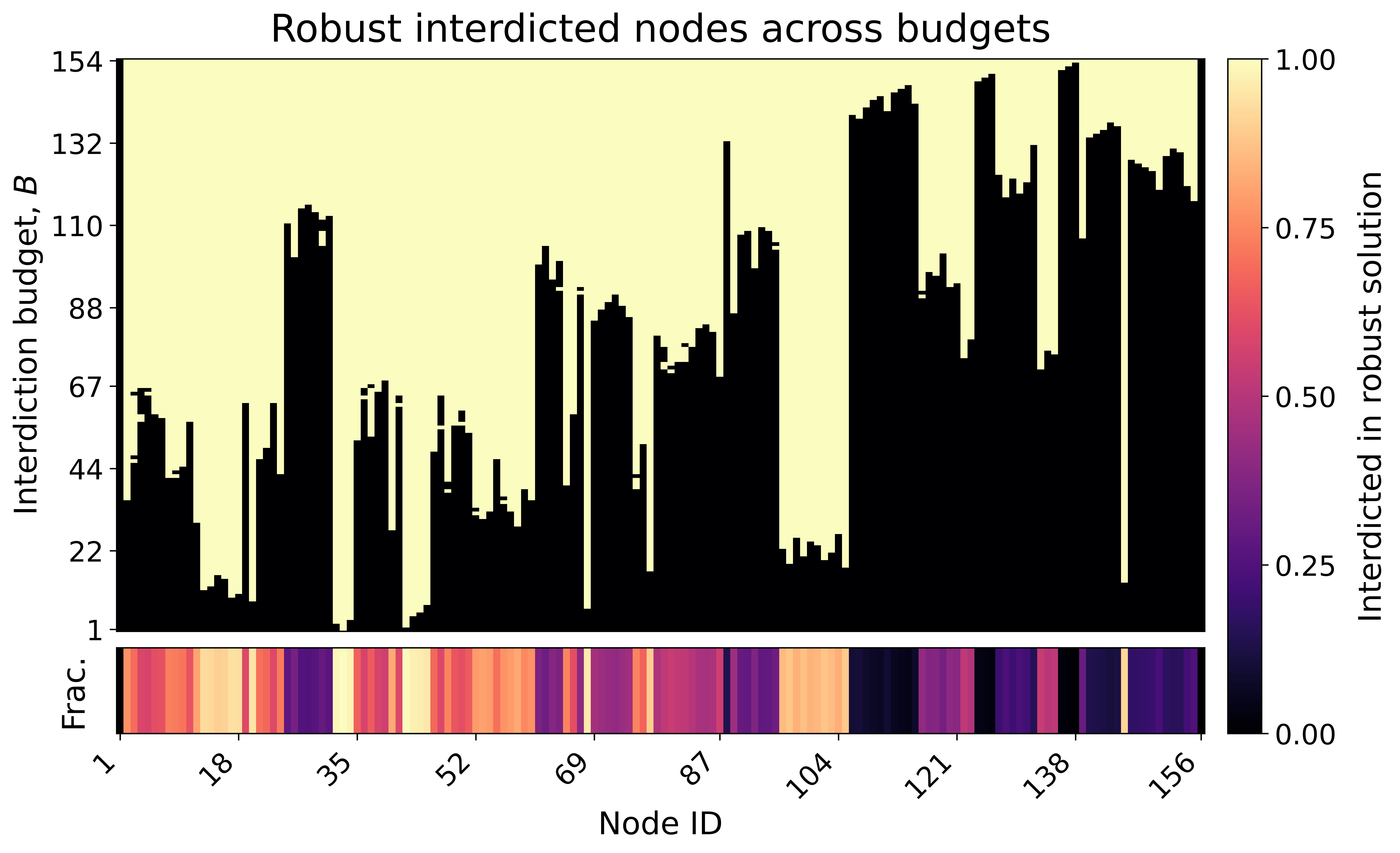}

    \includegraphics[width=0.42\linewidth, clip, trim= 0 2cm 0 0]{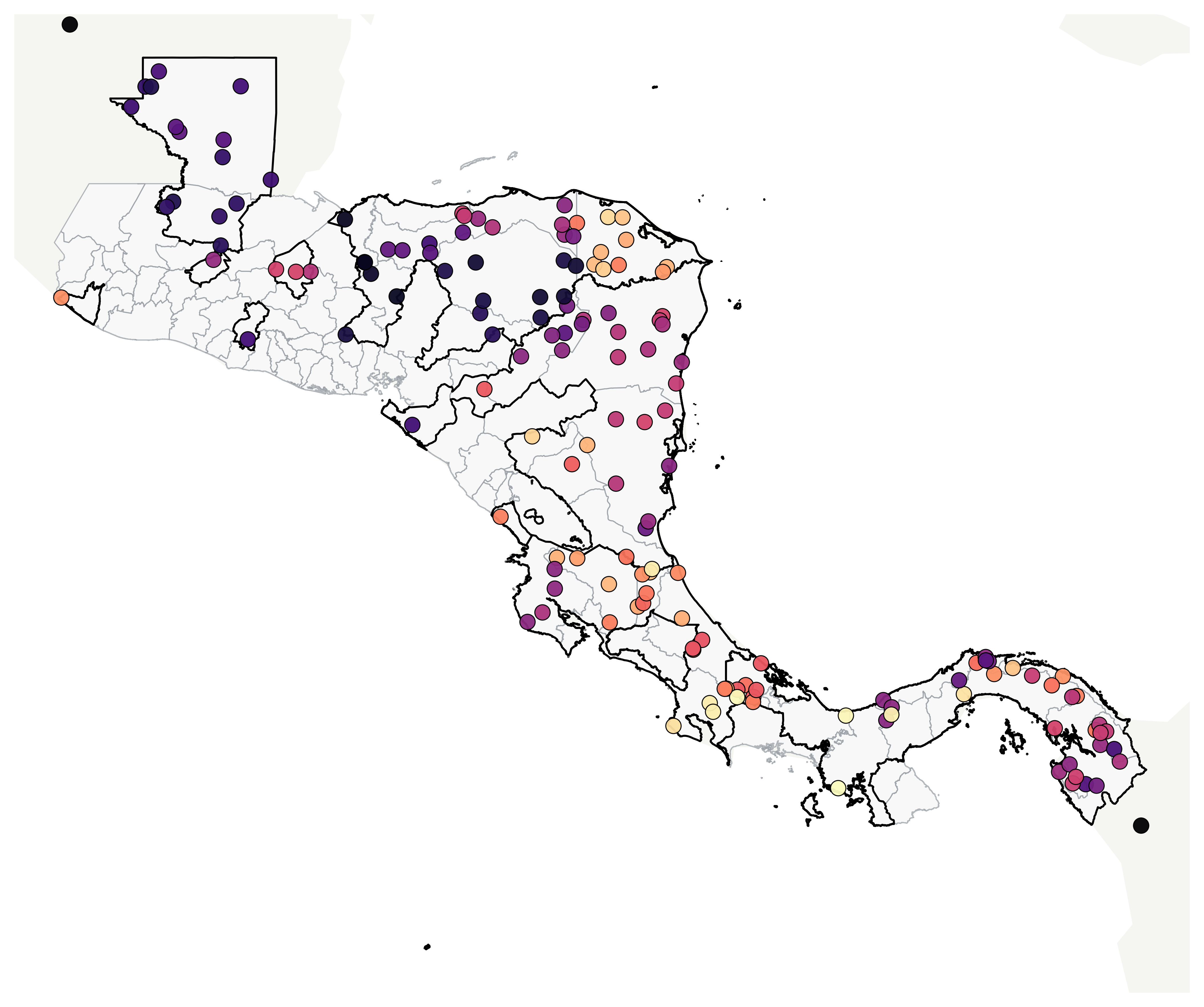}
    \includegraphics[width=0.49\linewidth]{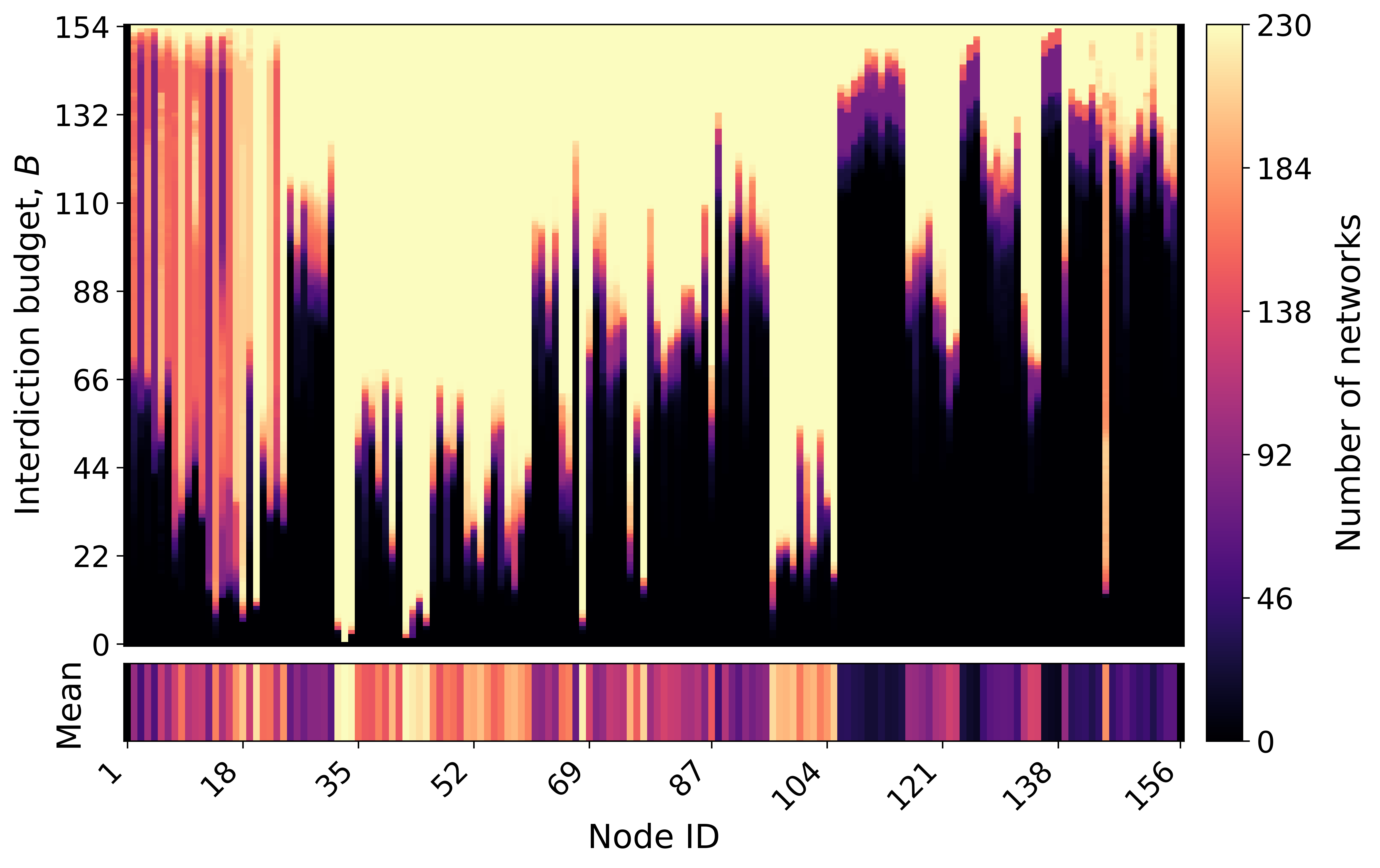}

    \caption{\textbf{Distribution of robust and optimal interdicted nodes over the filtered ensemble of networks.}
    (Left) Geographic visualization of the mean interdiction occurrence of each node across all budgets and data-consistent network realizations. (Right) Heat map of interdiction frequency across budgets and networks. The bottom strip shows the mean occurrence of each node across all budgets.}
\label{fig:interdiction_distribution}
\end{figure}

\begin{figure}[t!]
    \centering
    \includegraphics[width=0.90\linewidth, trim=0cm 13cm 0cm 8cm, clip]{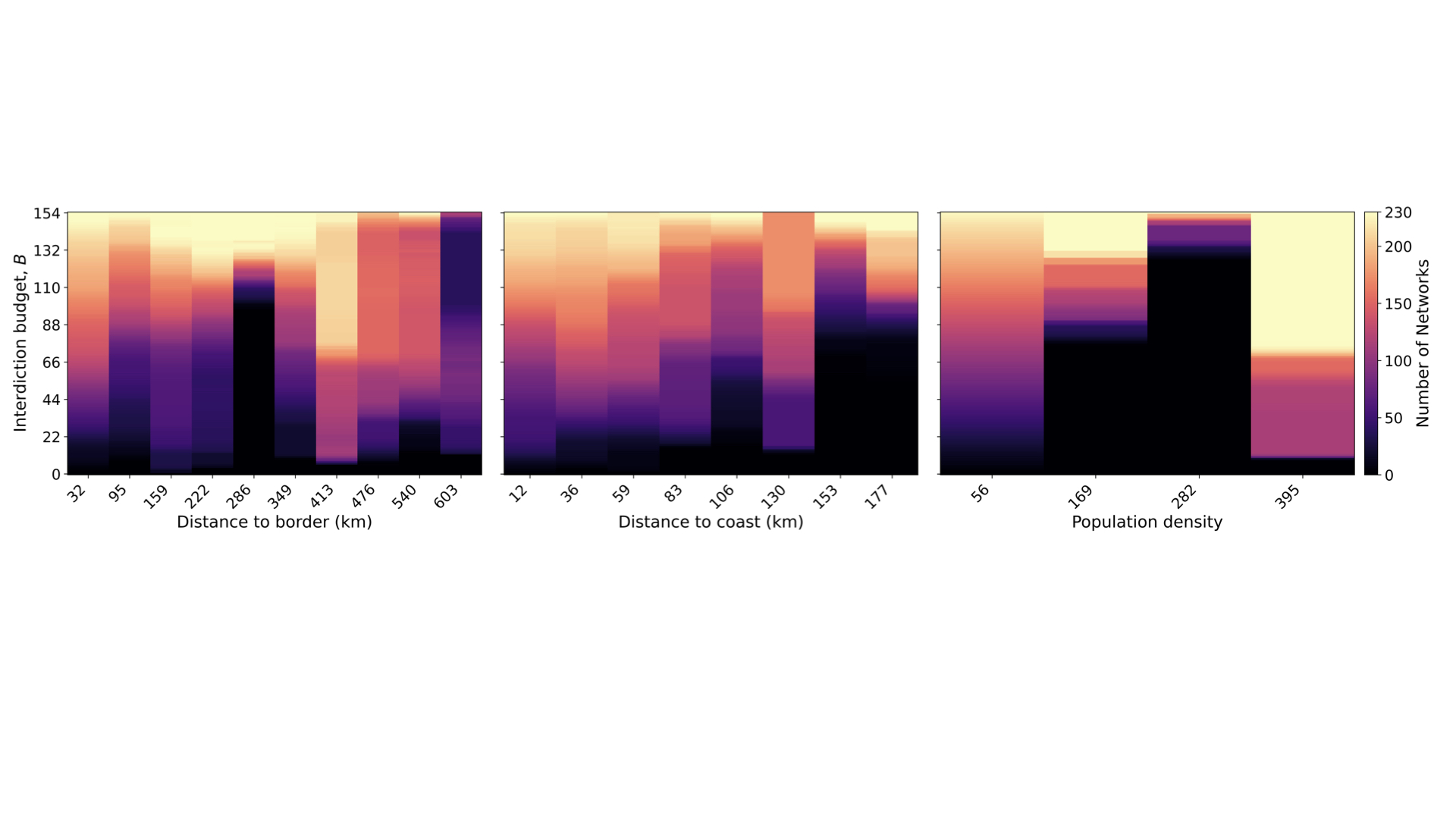}

    \includegraphics[width=0.90\linewidth]{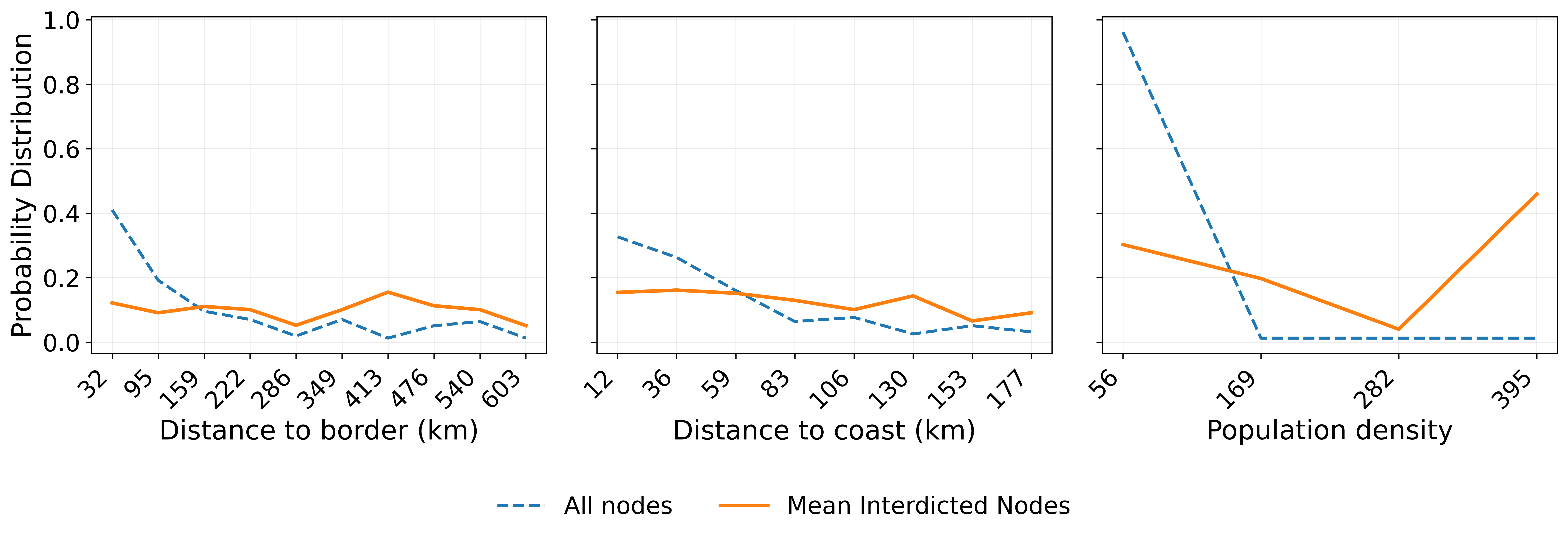}
    \caption{Feature Distribution of optimal interdicted nodes across the filtered ensemble of networks.}
    \label{fig:heatmap_distribution_optimal_feature}
\end{figure}

\begin{figure}[b!]
\centering
\includegraphics[width = 0.99\linewidth]{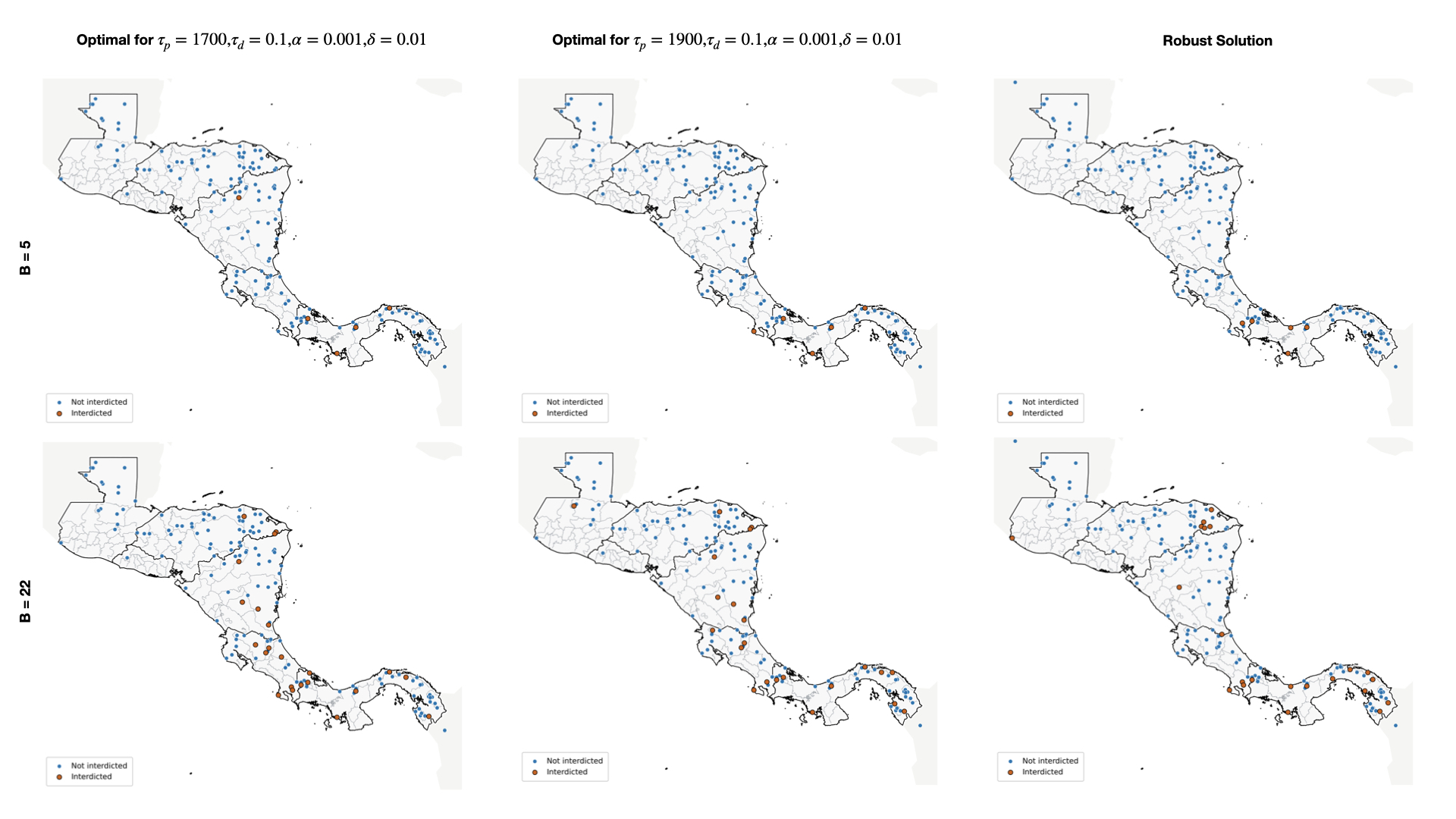}

\caption{Interdiction results for budgets $B=5$ and $B=22$. We present the optimal interdiction decision for two representative networks from the set of filtered ensemble of networks. Additionally, we also visualize the robust interdiction decision over the entire filtered ensemble  of networks. We observe that the set of interdicted nodes varies across the board, making the interdiction dependent on the graph topology and capacity.}
\label{fig:NodeBudget_interdiction_tau}

\end{figure}

\clearpage
\section{Sensitivity of Robust Interdiction to Ensemble of Networks}

\begin{figure}[h!]
    \centering
    \includegraphics[width=0.49\linewidth]{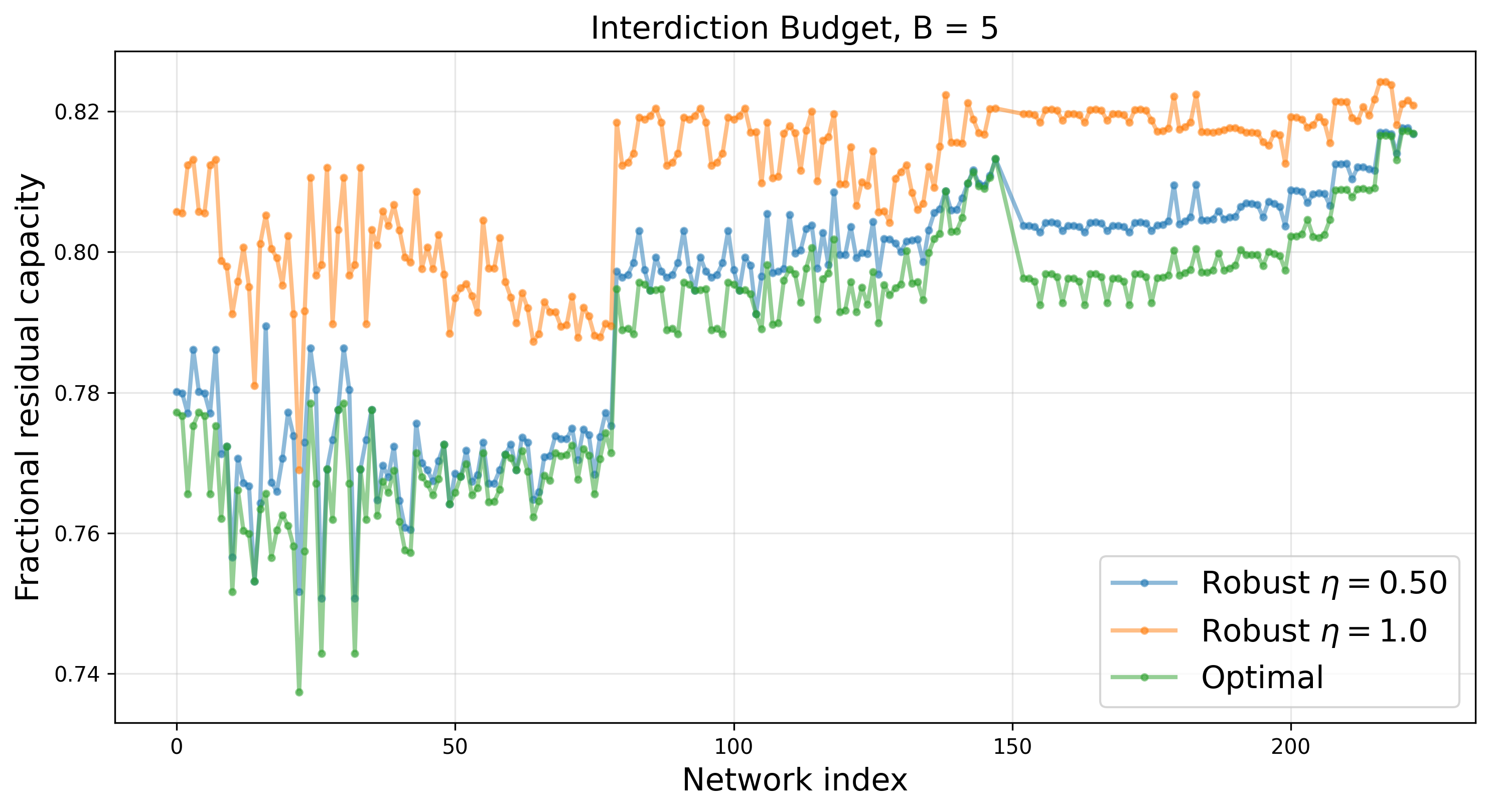}
    \includegraphics[width=0.49\linewidth]{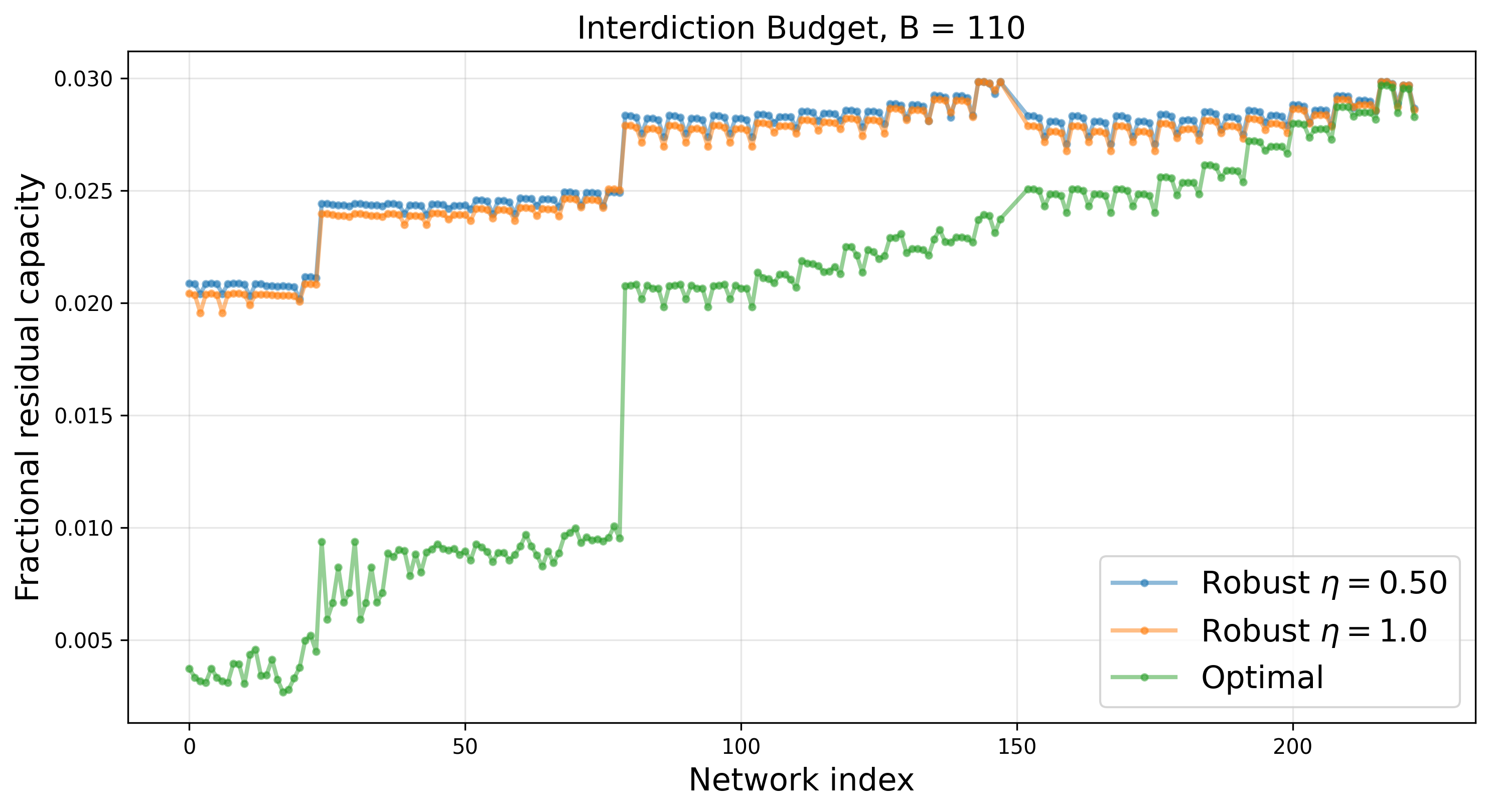}
    \caption{\textbf{Sensitivity analysis of Robust Interdiction to ensemble filtering threshold.} We illustrate the fractional residual capacity in the filtered ensemble of networks at $\eta = 0.50$, which are referenced by the network index for each of the 230 networks. For each network, we plot the resulting capacity for the robust solution obtained over ensemble of networks filtered at $\eta$, given interdiction budget $B$.}
    \label{fig:dist_ensemble_filtering}
\end{figure}

\paragraph{Setup.} We next examine how the choice of ensemble filtering threshold $\eta$ affects the quality and computational cost of robust interdiction. Table~\ref{tab:robust_ilp_sizes} reports the resulting network, scenario, and ILP sizes as $\eta$ varies. Increasing the threshold admits more networks into the robust ensemble, which substantially increases the number of scenarios, variables, and constraints. Figure~\ref{fig:dist_ensemble_filtering} illustrates the robust interdiction solutions obtained under different ensemble thresholds on the filtered ensemble at $\eta=0.50$, using the network index to compare fractional residual capacity across networks.

\begin{table*}[h]
\centering
\footnotesize
\caption{Network, scenario, and ILP sizes for robust interdiction experiments under different ensemble thresholds $\eta$. The number of scenarios and ILP size vary with the ensemble threshold. $\eta = $``None" refers to the full ensemble of networks generated using simulation scenarios from Table~\ref{tab:network_generation_parameters}.}
\label{tab:robust_ilp_sizes}
\resizebox{\textwidth}{!}{%
\begin{tabular}{lrrrrrrrrrrrr}
\toprule
$\eta$ 
& $|V|$ 
& Scenarios 
& Total \# edges  
& Mean \# edges 
& \# Variables 
& \# Constraints \\
\midrule
0.50 
& 156  
& 230 
& 268,942  
& 1,169.31 
& 573,921 
& 1,076,461\\


1.00 
& 156
& 865 
& 1,091,887 
& 1,262.30 
& 2,318,871 
& 4,370,146 \\

None 
& 156 
& 2,560 
& 3,039,584 
& 1,187.34 
& 6,478,685 
& 12,166,019\\
\bottomrule
\end{tabular}%
}
\vspace{0.5em}
\end{table*}

The results show that ensemble filtering has the largest effect at lower interdiction budgets. When $B=5$, the robust solutions obtained under different thresholds lead to visibly different residual capacities, indicating that the selected ensemble strongly influences which interdiction strategy is most effective. In contrast, at higher budgets such as $B=110$, the robust solutions obtained from different ensemble thresholds are much closer in quality. Thus, although larger ensembles lead to substantially larger ILPs, the improvement in solution quality becomes nominal at high budgets. Overall, these results suggest that filtering provides a useful tradeoff: it reduces computational cost significantly while preserving much of the robust-solution quality, especially when the marginal gain from adding more scenarios is small.

\end{document}